\documentclass[useAMS,usenatbib]{mnras}
 \usepackage{dblfloatfix} 
 \usepackage{epsfig, color} \usepackage{graphicx}
 \usepackage[fleqn]{amsmath} \usepackage{epstopdf}
 \usepackage{caption} \usepackage{float}

 \def\ltsima{$\;\buildrel < \over \sim \;$} 
 \def\simlt{\lower.5ex\hbox{\ltsima}}
 \def\gtsima{$\; \buildrel >\over\sim\;$}
 \def\simgt{\lower.5ex\hbox{\gtsima}}

\title[AGN jet heating of realistic clusters]{AGN jet feedback on a moving mesh: gentle cluster heating by weak shocks and lobe disruption} \author[]{Martin A.
  Bourne$^{1,2,\star}$ and Debora Sijacki$^{1,2}$\\ 
  $^{1}$ Institute of Astronomy, University of Cambridge, Madingley Road, Cambridge, CB3 0HA, UK\\ 
  $^{2}$ Kavli Institute for Cosmology, University of Cambridge, Madingley Road, Cambridge, CB3 0HA, UK\\ 
  $^{\star}$ {E-mail:~} {\rm mab231@cantab.ac.uk} }

\begin{document} 
\date{Received} \pagerange{\pageref{firstpage}--\pageref{lastpage}}
\pubyear{2020} \maketitle \label{firstpage}

\maketitle

\begin{abstract} 
While there is overwhelming observational evidence of AGN-driven jets in galaxy clusters and groups, 
{\it if} and {\it how} the jet energy is delivered to the ambient medium remains unanswered. Here we perform very high resolution AGN jet simulations within a live, cosmologically evolved cluster with the moving mesh code {\sc arepo}. We find that mock X-ray and radio lobe properties are in good agreement with observations with different power jets transitioning from FR-I to FR-II-like morphologies. During the lobe inflation phase, heating by both internal and bow shocks contributes to lobe energetics, and $\sim 40$ per cent of the feedback energy goes into the $PdV$ work done by the expanding lobes. Low power jets are more likely to simply displace gas during lobe inflation, but higher power jets become more effective at driving shocks and heating the intracluster medium (ICM), although shocks rarely exceed $\mathcal{M}\sim 2-3$. Once the lobe inflation phase ceases, cluster weather significantly impacts the lobe evolution. Lower power jet lobes are more readily disrupted and mixed with the ICM, depositing up to $\sim 70$ per cent of the injected energy, however, ultimately the equivalent of $\simgt 50$ per cent of the feedback energy ends up as potential energy of the system. Even though the mean ICM entropy is increased up to $80$~Myr after the jets switch off, AGN heating is gentle, inducing no large variations in cluster radial profiles in accord with observations. 
\end{abstract}

\begin{keywords} galaxies: active, jets - galaxies: clusters: general, intracluster medium - black hole physics - methods: numerical \end{keywords}

\section{Introduction}
\label{sec:intro}
Galaxy clusters, which due to the nature of hierarchical structure formation, build up through successive mergers of smaller systems over cosmic time to become the most massive, gravitationally bound systems in the Universe. Comprising of a massive dark matter halo, menagerie of galaxies and the hot intracluster medium (ICM), they represent a rich ecosystem in which to study a range of astrophysical phenomena, as well as being complimentary cosmological probes \citep[see e.g.,][for reviews]{AllenEtAl11, KravtsovBorgani12, McNamaraNulsen12}. The ICM, which radiates energy through thermal Bremsstrahlung, provides a key channel through which clusters can be observed in the X-ray band, with its exact properties being determined by a combination of astrophysical processes and cosmology. It was originally thought that left unabated, high central radiative losses would result in the formation of a cooling flow \citep{Fabian94}, leading to large quantities of cold gas dropping out of the ICM and hence massive molecular reservoirs and high star formation rates (SFRs) within cluster centres. However, cluster cores typically show only moderate molecular gas content \citep{fogarty2019dust,castignani2020molecular,McNamaraEtAl14,RussellEtAl14,RussellEtAl17,russell2019driving} and low SFRs \citep{McNamara2007,DonahueEtAl10,CookeEtAl2016,FogartyEtAl17}, although there can be exceptions \citep[e.g.,][]{CrawfordEtAl99,EgamiEtAl06,von2007special,MittalEtAl15,FogartyEtAl17}. Additionally, direct observational signatures of the cooling, such as X-ray emission lines below $1$~keV are absent from numerous observations \citep[e.g.,][]{IkebeEtAl97,MakishimaEtAl01,PetersonEtAl01,PetersonEtAl03,TamuraEtAl01,BoehringerEtAl2002,MatsushitaEtAl02,LewisEtAl2002} and UV detections of O VI lines suggest lower than expected cooling rates \citep[e.g.,][]{OegerleEtAl01,BregmanEtAl06,donahue2017observations}. 

 Instead, heating by active galactic nucleii (AGN) feedback, possibly in combination with other physical processes \citep[for example thermal conduction,][]{RuszkowskiBegelman02, ZakamskaNarayan03, VoigtFabian04,ConroyOstriker08,BogdanovicEtAl09,RuszkowskiOh10,RuszkowskiOh11,YangReynolds16ThermalConduction}, is very likely to be able to keep the ICM warm and prevent the formation of the aforementioned cooling flows \citep[see e.g.][for reviews]{Fabian12,McNamara2007,McNamaraNulsen12}. Specifically, accretion onto the central black hole (BH) can produce jets that inflate lobes of relativistic gas in cluster cores \citep{BinneyTabor95,OmmaEtAl04,McNamaraEtAl05,FabianEtAl06,SijackiEtAl06a,SijackiEtAl07,CattaneoEtAl07,FormanEtAl07,DuboisEtAl10,DuboisEtAl12}, which exhibit themselves as cavities in X-ray observations \citep{FabianEtAl2000, FabianEtAl2011, McNamaraEtAl2000, HeinzEtAl02, FormanEtAl07}. Along with radio emission, such cavities (or bubbles) are found to be common in cool core clusters, i.e. those with short central cooling times \citep{Burns90,DunnEtAl05,McNamara2007,DunnFabian06,DunnFabian08,Sun09,Fabian12,Hlavacek-LarrondoEtAl12}. Combining this with the observed correlation between the energy content of the lobes (based on $PV$ estimates) and the ICM cooling losses \citep[based on X-ray luminosity, e.g.,][]{RaffertyEtAl06,McNamara2007,NulsenEtAl07,DunnFabian08,Fabian12}, strongly suggests that ICM cooling and heating is coupled to the AGN activity.  
 
 While the feedback energetics are sufficient to offset the expected ICM cooling losses, and despite a growing consensus that jet feedback in galaxy clusters should be gentle and continuous \citep[see discussions in][for examples]{McNamara2007,McNamaraNulsen12,Fabian12,WernerEtAl19}, there is still some debate with regard to the mechanism (or combination thereof) through which the mechanical energy of the jets is effectively and (largely) isotropically communicated to the ICM. Perhaps the clearest heating mechanism seen in observations is via the weak shocks driven into the ICM by the lobe inflation \citep{FabianEtAl03,RandallEtAl15} that are also seen in simulations of jet feedback \citep{LiEtAl2016, YangReynolds16Hydro, BourneSijacki17, MartizziEtAl18}. Such shocks should slow as they propagate into the ICM and broaden into sound waves that could transport energy to larger radii, with some works interpreting observed ripples in X-ray brightness maps as sounds waves \citep{FabianEtAl03,FabianEtAl05,FabianEtAl17}. Although simulations also exhibit sound wave production \citep{RuszkowskiEtAl04,SijackiEtAl06a,BourneSijacki17,BambicEtAl19}, the volume over and rate at which they heat the ICM will depend upon the ICM viscosity set on micro-physical scales \citep{RuszkowskiEtAl04, SijackiEtAl06b, ZweibelEtAl18}. Other mechanisms provide less direct or obvious observational signatures; such as instability driven mixing \citep{HillelSoker16,HillelSoker17}, cavity heating \citep{ChurazovEtAl02, BirzanEtAl04}, jet driven turbulence \citep{ZhuravlevaEtAl2014, ZhangEtAl18}, cosmic ray (CR) production \citep{SijackiEtAl08, Pfrommer13,EhlertEtal18} and gas circulation \citep{YangReynolds16Hydro}. 
 
Given that a number of these processes do not leave a directly or easily observable signature, the role of simulations is critical to determine how they manifest themselves in cluster observables and thus ultimately resolve what processes dominate the heating budget. Previous simulation works typically sit within one of two categories. On the one hand, cosmological simulations self-consistently capture the build up and evolution of galaxy clusters over cosmic time and hence by construction exhibit the cluster properties and dynamics expected from large scale structure formation theory \citep[e.g.][]{BaheEtAl17, BarnesEtAl17,BarnesEtAl17b,BarnesEtAl18,McCarthyEtAl17,RasiaEtAl15,HahnEtAl17,HendenEtAl18}. While, such simulations provide realistic cluster environments, their large spatial and temporal dynamic range means that they typically lack sufficient resolution in cluster cores, close the to central BH, to include detailed modelling of the AGN feedback process. On the other hand, previous works including physically more realistic models of jet injection that self-consistently capture the lobe inflation and its interaction with the ICM are restricted in their dynamic range by the requirement of high resolution within the jets and lobes and hence adopt idealised ICM setups \citep[e.g.,][]{BourneSijacki17,HardcastleKrause13,HardcastleKrause14,EnglishEtAl16,Krause03,OmmaEtAl04,WeinbergerEtAl17,VernaleoReynolds06,ReynoldsEtAl01,BassonAlexander03,GaiblerEtAl09}. Up until recently, only a handful of examples of an AGN jet feedback model included in a cosmological environment exist in the literature \citep{HeinzEtAl06, MorsonyEtAl2010, MendygralEtAl12}. Our previous study, \citet{BourneEtAl19}, builds upon these early works by combining very high resolution jets, and uniquely a dedicated refinement scheme to ensure continued high resolution in the jet lobes and at the lobe-ICM interface, within a fully cosmological cluster that was evolved with state-of-the-art physical sub-grid models. Here we further expand on this work by including jets of different power, and perform a detailed analysis of the lobe evolution and cluster heating processes with the aim to address the crucial issues of {\it how the jet energy is communicated to the ICM} and {\it if it is sufficient to hinder cooling flows}.

The paper is structured as follows. In Section~\ref{sec:method} we outline the numerical implementation and initial condition (IC) generation and properties. Section~\ref{sec:results} presents our main results and is split into sections covering lobe evolution (\ref{sec:lobe_evo}), how energy is communicated to the ICM (\ref{sec:icm_heating}) and the impact on cluster properties (\ref{sec:cluster_properties}). Finally, in Section~\ref{sec:discussion} we discuss our results and limitations, and in Section~\ref{sec:summary} we provide a short summary of our key results.

\section{Numerical Method}
\label{sec:method}
\subsection{Simulation overview}    
\label{sec:sim_overview}
\begin{table*}  
\centering
\begin{tabular}{|l|c|c|c|c|c|c|c|c|c|c|c|}
\hline
Run & $\dot{E}_{\rm J}$/[erg/s] & $\dot{m}_{\rm Edd}$ & $v_{\rm J}^{\rm peak}$/[c] & $v_{\rm J}^{\rm mean}$/[c] & $\dot{M}_{\rm J}^{\rm trace}$/[$M_{\odot}$/yr] & $\chi_{\rm M}$ & $\tau_{\rm J}$/[Myr] & $f_{\rm J}^{\rm thresh}$ & Refinement criteria \\
\hline
low & $7.8\times 10^{43}$ & $4\times 10^{-5}$ & $0.29$ & $0.05$ & $\sim 0.3$ & $\sim 43$ & $20$ & $10^{-2.0}$ & SLR, JR, JFR, VR, GD \\
fiducial & $3.9\times 10^{44}$ & $2\times 10^{-4}$ & $0.47$ & $0.14$ & $\sim 0.44$ & $\sim 13$ & $20$ & $10^{-2.5}$ & SLR, JR, JFR, VR, GD \\
high & $1.94\times 10^{45}$ & $1\times 10^{-3}$ & $0.56$ & $0.26$ & $\sim 0.7$ & $\sim 4$ & $20$ & $10^{-3.0}$ & SLR, JR, JFR, VR, GD \\
no jet & - & - & - & - & - & - & - & - & SLR, JR, VR, GD \\
\hline
\end{tabular}
\caption{Summary of the simulations performed, where columns show (1) the run name, (2) the mean jet power over the course of the jet life time, (3) the assumed Eddington ratio, (4) the maximum velocity achieved in the jet, (5) the mean outward jet velocity calculated over material with $f_{\rm J}>0.9$ and at $t\simgt 5$~Myr, (6) the typical jet tracer ``injection'' rate at $t\simgt 5$~Myr, (7) the mass loading within the jet cylinder ($\chi_{\rm M}=\dot{M}_{\rm J}^{\rm trace}/\dot{M}_{J}$), (8) the jet lifetime, (9) the threshold jet tracer value for defining lobe material and (10) the various refinement criteria used during the simulation.}
\label{tab:jet_runs}
\end{table*}  

Simulations are performed using the moving mesh-code {\sc arepo} \citep{SpringelArepo2010}, which employs a finite-volume approach to solve for the hydrodynamics on an unstructured Voronoi mesh that moves with the flow. Our ICs are identical to those used in our previous work, \cite{BourneEtAl19}, and comprise of a cosmological zoom simulation of a galaxy cluster evolved from redshift $z=127$ down to $z\simeq 0.1$ using sub grid models almost identical to those employed for the original Illustris project \citep{GenelEtAl14,VogelsbergerEtAl14, SijackiEtAl2015}, bar a small difference in the radio mode AGN feedback model, and we point the reader to Section~\ref{sec:cluster_ICS} for a more detailed account the generation of the ICs, along with the initial properties of the cluster.

Subsequently, instead of using the radio feedback model that was employed up until $z=0.1$ \citep{SijackiEtAl07, SijackiEtAl2015} where AGN lobes are not inflated self-consistently, we activate the jet feedback model presented in \cite{BourneSijacki17} and discuss in Section~\ref{sec:jet_model}, to study a range of different jet powers. Building on \cite{BourneEtAl19}, as well as a {\it fiducial} jet power of $3.9\times 10^{44}$~erg~s$^{-1}$, we perform three additional simulations with {\it low} ($7.8\times 10^{43}$~erg~s$^{-1}$) and {\it high} ($1.94\times 10^{45}$~erg~s$^{-1}$) power jets, and one with no jets for comparison. All jets are active for $20$ Myr and their properties are summarised in Table~\ref{tab:jet_runs}. We follow the subsequent jet lobe evolution for a further $\sim 80$~Myr. The jet model relies on the super-Lagrangian refinement technique of \cite{CurtisSijacki15} combined with additional modifications, which are discussed in Section~\ref{sec:refinement_scheme}. We note that only the most massive BH in the cluster of interest is active during our simulations, and unlike the original simulations used to produce the ICs, for simplicity we do not include the effects of an AGN radiation field in the heating and cooling prescription. Additionally, while during the generation of the cluster ICs, we followed the supernovae wind model used in the Illustris simulations, where the mass loading, $\eta_{\rm w}=\dot{M}_{\rm w}/\dot{M}_{\rm *}$, is a function of the halo 1D velocity dispersion \citep[see][for model details]{VogelsbergerEtAl13}, for simplicity we instead use a fixed value of $\eta_{\rm w}=1$ for the post $z=0.1$ simulations presented here. All other sub-grid models and parameters remain unchanged.

\subsection{Jet feedback model}
\label{sec:jet_model}

The feedback model used in this study is identical to that used in \citet{BourneEtAl19} and is based heavily on the framework outlined in our previous work, \cite{BourneSijacki17}\footnote{Equations~(\ref{eq:mdot_jet})-(\ref{eq:pdot_jet}) presented here contain some small differences compared to those presented in Sections 2.2.1 and 2.2.2 of \citet{BourneSijacki17}. Specifically, our original equations were not fully self-consistent with the physical picture of radiatively driven jets. Therefore, we corrected for this by re-defining the jet mass loading, $\eta_{\rm J}$, and BH growth rate, $\dot{M}_{\rm bh}$, as well applying small modifications to the equations describing the jet mass, power and momentum (namely removing an erroneous factor $1-\epsilon_{\rm r}$). Additionally, we note that $\dot{M}_{\rm in}$ used here is equivalent to $\dot{M}_{\rm a}$ in \citet{BourneSijacki17}. In any case, our previous results remain unaffected by this inconsistency in our framework given that we assume jets of a fixed power.}, and that derived by \citet{OstrikerEtAl10a}. Here we give a brief overview and highlight any changes made to the model for this study. The BH accretion and jet properties are defined as follows. First, we assume that gas flows towards the central BH at a rate $\dot{M}_{\rm in}$, which, for simplicity we assume is fixed at a given Eddington fraction $\dot{m}_{\rm Edd}=\dot{M}_{\rm in}/\dot{M}_{\rm Edd}$ for each run, where the Eddington rate is calculated as
\begin{equation}
\label{eq:Eddington}
\dot{M}_{\rm Edd} = \frac{4\pi GM_{\rm bh}m_{\rm p}}{\epsilon_{\rm r}\sigma_{\rm T}c}\, ,
\end{equation}
$G$ is the gravitational constant, $M_{\rm bh}$ is the BH mass, $m_{\rm p}$ is the proton rest mass, $\sigma_{\rm T}$ is the Thompson scattering cross section, and $c$ is the speed of light. Subsequently we assume that a fraction, $\left(1+\eta_{\rm J}\right)^{-1}$, of this material feeds a BH accretion disc, while the rest is launched in radiatively driven jets at a rate of 
\begin{equation}
\dot{M}_{\rm J}=\frac{\eta_{\rm J}}{1+\eta_{\rm J}}\dot{M}_{\rm in}\, ,
\label{eq:mdot_jet}
\end{equation}
where
\begin{equation}
\label{eq:mass_loading}
\eta_{\rm J}=\frac{\dot{M}_{\rm J}}{\dot{M}_{\rm in}-\dot{M}_{\rm J}}=1\, ,
\end{equation}
is the assumed mass loading factor defined as the ratio between material entering the jets and material entering the accretion disc. Subsequently, the BH growth rate and jet power can be calculated as
\begin{equation}
\label{eq:mdot_bh}
\dot{M}_{\rm bh}=\left(1-\epsilon_{\rm r}\right)\left[\dot{M}_{\rm in}-\dot{M}_{\rm J}\right]=\frac{1-\epsilon_{\rm r}}{1+\eta_{\rm J}}\dot{M}_{\rm in}\, 
\end{equation}
and
\begin{equation}
\label{eq:edot_jet}
\dot{E}_{\rm J}=\epsilon_{\rm J}\epsilon_{\rm r}\left[\dot{M}_{\rm in}-\dot{M}_{\rm J}\right]c^{2}=\frac{\epsilon_{\rm J}\epsilon_{\rm r}}{1+\eta_{\rm J}}\dot{M}_{\rm in}c^{2}\, ,
\end{equation}
respectively, where $\epsilon_{\rm r}=0.2$ is the assumed radiative efficiency of accretion and $\epsilon_{\rm J}=1$ is the assumed coupling efficiency of the jet energy to local gas. For completeness, the velocity and resulting outward momentum flux of the sub-resolution jets would be
\begin{equation}
    \label{eq:v_jet}
    v_{\rm J}^{\rm sub-res}=\left(\frac{2\dot{E}_{\rm J}}{\dot{M}_{\rm J}}\right)^{1/2}=\left(\frac{2\epsilon_{\rm J}\epsilon_{\rm r}}{\eta_{\rm J}}\right)^{1/2}c\,
\end{equation}
and
\begin{equation}
\label{eq:pdot_jet}
\dot{p}_{\rm J}=\dot{M}_{\rm J}v_{\rm J}^{\rm sub-res}=\eta_{\rm J}\frac{\epsilon_{\rm J}\epsilon_{\rm r}}{1+\eta_{\rm J}}\left(\frac{2\epsilon_{\rm J}\epsilon_{\rm r}}{\eta_{\rm J}}\right)^{1/2}\dot{M}_{\rm in}c\, ,
\end{equation}
respectively. However, as we discuss below, due to mass loading in the jet cylinder and the fact that we employ the ``kinetic energy conserving'' injection scheme these momenta and velocities do not reflect those achieved on resolvable scales in the simulations.

As in \cite{BourneSijacki17}, the jet injection region is defined by a cylinder of constant opening angle such that the ratio of radius to half cylinder height is set to $3/2$. Unlike our previous work we apply an additional criterion for the minimum number of cells in each half cylinder such that the volume is minimised for the conditions $n_{\rm cell}^{\rm n/s}\geq 10$ and $M_{\rm cyl}\geq 10^{4}$ $h^{-1}$M$_{\odot}$, where $n_{\rm cell}^{\rm n/s}$ is the number of cells within the northern/southern half of the cylinder (these are not necessarily the same) and $M_{\rm cyl}$ is the total mass of cells within the whole cylinder. The jets are then injected equally into each half cylinder, directed in opposite directions along the $z$-axis. Cylinder gas cell properties are updated by injecting mass, momentum and internal energy into the cells weighted by the kernel function
\begin{equation}
W_{\rm i}(r,z)=\frac{V_{\rm i}\exp\left(-\frac{r^{2}}{2r_{\rm cyl}^{2}}\right)|z|}{V_{\rm weight}}\, ,
\end{equation}
where $V_{\rm i}$ is the cell volume, $V_{\rm weight}$ is the kernel function normalisation factor such that $\sum_{\rm i}{W_{\rm i}(r,z)}=1$, and $r=\sqrt{x^2+y^2}$ and $z$ are cylindrical co-ordinates of the $i^{\rm th}$ cell, respectively. The weighting is performed separately for each half cylinder, such that during a timestep $dt$, the total mass and energy injected in to each half cylinder is equal to $\dot{M}_{\rm J}\times dt/2$ and $\dot{E}_{\rm J}\times dt/2$, and we note that an ideal equation of state with $\gamma =5/3$ is used for all gas in our simulations (see Section~\ref{sec:discussion} for further discussion).

By injecting the jets into a cylinder of pre-existing material, they are by construction mass loaded, resulting in two ways in which the jet momentum and energy can be injected \citep{BourneSijacki17}. One could explicitly conserve the outward momentum flux\footnote{Strictly speaking, provided equal momentum is given to both jets, the symmetry of the injection means that net momentum is always conserved and equal to zero.} and update cell momenta such that Equation~(\ref{eq:pdot_jet}) is conserved. This would lead to the resulting jet kinetic energy falling short of that required by Equation~(\ref{eq:edot_jet}), although additional thermal energy could be added to jet cells \citep[e.g.,][]{CattaneoEtAl07,OmmaEtAl04} to account for this. Alternatively, one can inject momentum such that the resulting kinetic energy of the jet satisfies Equation~(\ref{eq:edot_jet}) \citep[e.g.,][]{DuboisEtAl10,GaspariEtAl11,GaspariEtAl12,YangReynolds16Hydro}, however, in this case the total outward momentum flux would exceed that given by Equation~(\ref{eq:pdot_jet}). Following previous works \citep[e.g.,][]{GaspariEtAl11,GaspariEtAl12,PrasadEtAl15,YangReynolds16Hydro,YangReynolds16ThermalConduction} that successfully reproduce features of cool-core (CC) clusters in idealised setups, we adopt the latter approach. Cells with initial mass, momentum and kinetic energy of $m_{i,0}$, $p_{i,0}$ and $E_{i,0}$, are given a momentum kick along the jet direction of
\begin{equation}
    \label{eq:dpdot}
    |dp_{i}|=\sqrt{2(m_{i,0}+dm_{i})(E_{i,0}+dE_{i}^{tot})}-|p_{i,0}|,
\end{equation}
where $dm_{i}$ is the mass of jet material injected into the cell and $dE_{i}^{tot}$ is the desired total energy gain of the cell. The total injected energy might not be conserved due to momentum cancellation, therefore, if the actual gain in a cells total energy is less than $dE_{i}^{tot}$, we correct for this by increasing the cells internal energy by an appropriate amount (typically accounting for $\simlt 2\%$ of the injected energy). The amount of resulting mass loading ($\chi_{\rm M}=\dot{M}_{\rm J}^{\rm trace}/\dot{M}_{J}$) in the cylinders is shown for each run in Table~\ref{tab:jet_runs}, along with the rate of jet tracer production ($\dot{M}_{\rm J}^{\rm trace}$), which gives an indication of the actual rate at which mass is injected into the cluster core from the cylinder. As expected lower power jets undergo higher mass loading due to the higher ratio between cylinder and jet masses.

Two tracers are used to track jet material. The {\it jet tracer}, $f_{\rm J}=m_{\rm J}^{\rm trace}/m_{\rm cell}$, is defined for each cell as the fraction of mass within that cell that has been within the jet cylinder during an injection episode, i.e. we set $f_{\rm J}=1$ whenever a cell is in the jet cylinder and the jet is active. The tracer mass, $m_{\rm J}^{\rm trace}$, is subsequently advected between cells in line with the total gas mass. We define lobe material as any non-star forming gas ($n<0.26$ h$^2$ cm$^{-3}$) for which $f_{\rm J}>f_{\rm J}^{\rm thresh}$. In \cite{BourneEtAl19}, we noted that while there is some freedom in the choice of $f_{\rm J}^{\rm thresh}$, a cells internal energy should be dominated by jet material when $f_{\rm J}\geqslant T_{\rm ICM}/T_{\rm J}$, where $T_{\rm ICM}$ and $T_{\rm J}$ are the typical ICM and jet-rich lobe material temperatures, respectively. For our fiducial jet power $f_{\rm J}^{\rm thresh}=10^{-2.5}$ provides a threshold that coincides with the point at which gas density (temperature) starts to decrease (increase), transitioning from ICM to lobe like values. However, higher (lower) power jets result in clearly higher (lower) temperature gas by roughly a factor of a few. As such we adjust values of $f_{\rm J}^{\rm thresh}$ for different jet powers (see Table~\ref{tab:jet_runs}), noting that this is not an exact art and as such the values chosen reflect the approximate transition between jet and non-jet material. Additionally, we inject a second tracer that tracks the mass injected via Equation~(\ref{eq:mdot_jet}), $m_{\rm J}^{\rm inj}$, which like our primary tracer is advected between cells.

\begin{figure*}
\psfig{file=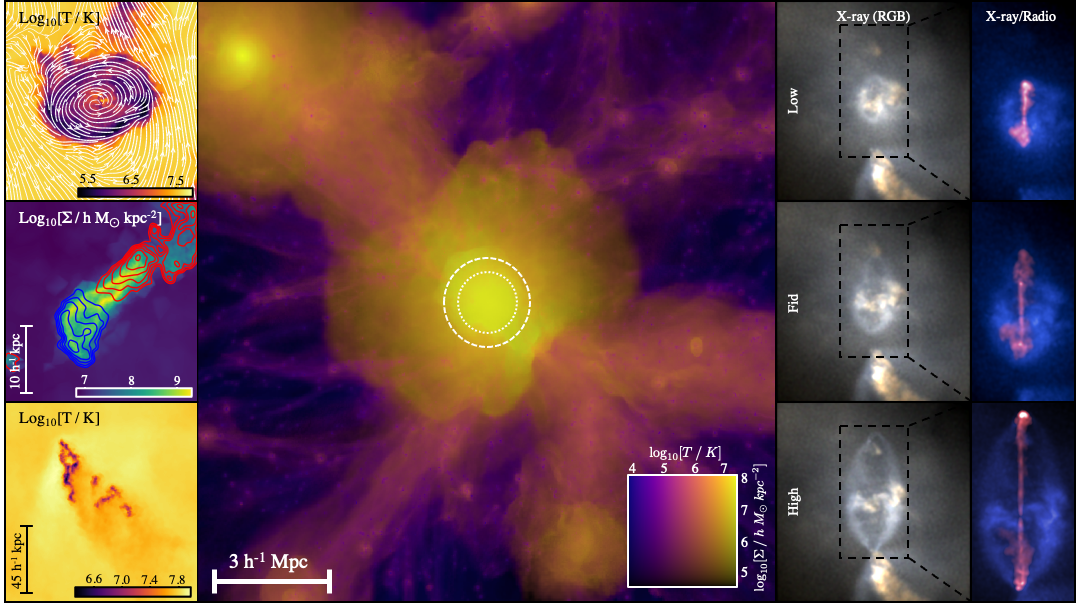,width=1.\textwidth}
\caption{Overview of cluster environment and jet evolution. The {\it large central panel} shows a column-density/temperature map of the main cluster, projected through a box of $15$~h$^{-1}$~Mpc on a side, with the colour indicating temperature and brightness the gas density. On the {\it left hand side} are three smaller stacked panels that show: a face-on temperature projection of the cold disc structure constituting the BCG at the centre of the cluster (top), side-on column density map of the same structure (middle), and a temperature projection of a cold substructure that falls into the cluster centre and interacts with the jet (bottom). The temperature projection of the BCG is overlaid with velocity stream-lines while the the density projection shows line-of-sight velocity contours, illustrating the rotation of the cold structure. To the right of the main panel is a grid of six mock-observations of different power jets (low to high power, from top to bottom) at $\sim 20$~Myr. The left hand column shows composite X-ray images, in which the different colours represent different energy bands and the right hand column shows a X-ray/radio composite images in which the blue shows the high energy X-ray band and the pink shows the synchrotron emissivity as a proxy for the radio band.}
\label{fig:overview}
\end{figure*}

\subsection{Refinement criteria}
\label{sec:refinement_scheme}

The standard refinement scheme within {\sc arepo} acts to ensure that all gas cells have a set target mass within some tolerance factor, however, the inclusion of the jet feedback scheme requires additional criteria that supersede this. The super-Lagrangian refinement (SLR) scheme of \cite{CurtisSijacki15} forces high resolution close to the BH. The refinement region is defined as a sphere of radius $h_{\rm bh}$ such that the mass of cells within $h_{\rm bh}$ is approximately $M(h_{\rm bh})\simeq n_{\rm ngb}^{\rm bh}\times m_{\rm cell}^{\rm target}$, where $n_{\rm ngb}^{\rm bh}=32$ is the number of neighbouring cells to the BH and $m_{\rm cell}^{\rm target}$ is the target gas cell mass. The SLR scheme ensures a smooth transition in cell radii from $r_{\rm cell}=h_{\rm bh}/2$ at the refinement region boundary to $r_{\rm cell}=r_{\rm Bondi}$ in the centre. Additionally, gentle de-refinement (GD) ensures that neighbouring cells can only merge if the gradients between their properties are sufficiently small, and we also only allow cells with $f_{\rm J} < 10^{-2}$ to de-refine. An additional jet refinement (JR) criteria is applied whereby cells within a distance of $2r_{\rm J}$ of the central BH are refined if their mass meets the criteria 
\begin{equation}
\frac{m_{\rm cell}}{M_{\rm cyl}}>(\alpha -\beta)\left(\frac{r}{r_{\rm J}}\right)^{\kappa}+\beta,
\end{equation}
where $\alpha =10^{-3}$, $\beta =10^{-2}$ and $\kappa =\left[\ln(1-\beta)-\ln(\alpha -\beta)\right]/\ln{2}$. This ensures that the jet cylinder is well populated while not becoming excessively massive or large. Further, in this work we impose that the jet lobes remain well resolved by implementing a jet fraction refinement criteria (JFR) whereby cells with $\log_{10}f_{\rm J} > -5$ are refined if their volume exceeds $V_{\rm min}^{\rm J}\times\left[1-\log_{10}f_{\rm J}\right]$, and  similar to \cite{WeinbergerEtAl17}, a volume refinement criteria (VR) is used to ensure a smooth transition between the sizes of neighbouring cells by limiting the maximum ratio of neighbouring cell volumes to no more than eight. To establish a high resolution jet injection region, refinement criteria are activated at the beginning of the simulation and the system is allowed to evolve for $\sim 6$~Myr, after which the jet is switched on, corresponding to $t=0$ in the following analysis.

\subsection{Gas draining}
\label{sec:gas_drain}
We drain gas at a rate of $\dot{M}_{\rm in}$ from certain cells within $h_{\rm bh}$ to ensure energy conservation, in a method similar to \citet{BourneSijacki17}, which is adapted from previous methods \citep{VogelsbergerEtAl13,CurtisSijacki15}. During a BH timestep of size $dt$, a mass of $M_{\rm drain}=\dot{M}_{\rm in}\times dt$ is removed from the cells that lie within a torus shaped region outside of the jet cylinder, aligned perpendicular to the jet direction and with an opening angle of $\left(\pi - \theta_{\rm J}\right)$, in a mass-weighted fashion. Subsequently, the BH mass is increased by $\Delta M_{\rm bh}=\dot{M}_{\rm bh}\times dt$ (see Equation~(\ref{eq:mdot_bh})) and a mass of $\Delta M_{\rm J}=\dot{M}_{\rm J}\times dt$ is added to the cells within the jet cylinder following the same weighting as the energy and momentum injection. Note that this method is designed to conserve energy, not mass, as a fraction $\epsilon_{\rm r}/(1+\eta_{\rm J})$ of $M_{\rm drain}$ is assumed to go into powering the jet during each timestep. 

\subsection{Cluster initial conditions}
\label{sec:cluster_ICS}

The ICs for this work are taken from a cosmological zoom simulations of a galaxy cluster evolved to low redshift, assuming the WMAP-9 cosmological parameters \citep{HinshawEtAl13} $\Omega_{\rm m}=0.2726$, $\Omega_{\Lambda}=0.7274$, $\Omega_{\rm b}=0.0456$, $\sigma_{8}=0.809$, $n_{\rm s}=0.963$ and $H_{0}=70.4$ km s$^{-1}$ Mpc$^{-1}=h\times 100$ km s$^{-1}$ Mpc$^{-1}$, as in the original Illustris simulations \citep{GenelEtAl14,VogelsbergerEtAl14}. The zoom-in region encompasses material that resides within a volume spanning $\sim 30$~h$^{-1}$~cMpc across. High resolution gas cells have a target mass, $m_{\rm cell}^{\rm target}=1.37\times~10^{7}$~h$^{-1}$~M$_{\odot}$ and adaptive softening lengths equal to $2.5\times r_{\rm cell}$, while dark matter particles have a mass of $m_{\rm dm}=4.83\times~10^{7}$~h$^{-1}$~M$_{\odot}$ with physical gravitational softening lengths set to $2.8125$~h$^{-1}$~kpc below $z=5$.

Sub-grid models for gas radiative processes \citep{KatzEtAl96, WiersmaEtAl09a}, interstellar medium  (ISM) and stellar physics \citep{Springel03,Springel05,VogelsbergerEtAl13,WiersmaEtAl09b}, and BH  processes \citep{DiMatteiEtAl05, Springel05, SijackiEtAl07} are almost  identical  to those employed in the original Illustris project. In producing the ICs, the radio-mode model of \cite{SijackiEtAl07} is employed during low Eddington rate phases of BH growth. The feedback acts by injecting hot bubbles of gas (to mimic radio lobes) around accreting BHs whenever a BH’s mass increases by a fraction $\delta_{\rm BH}$. For the model parameters chosen in the original Illustris project, namely $\delta_{\rm BH}=0.15$, too much gas was ejected from groups and clusters \citep{GenelEtAl14}. Therefore in this work we apply a gentler, albeit more frequent bubble injection by setting $\delta_{\rm BH}=0.015$, resulting in a gas mass fraction of $f_{\rm gas}^{500}=14\%$ within $R_{500,\:c}$ for our ICs. 
\begin{table}
  \centering
  \begin{tabular}{|l|c|}
    \hline
    \multicolumn{1}{}{} & Initial properties \\
    \hline
    $z$ & $0.0989$ \\
    $M_{\rm 500}$ & $2.82\times 10^{14}$ h$^{-1}$ M$_{\odot}$\\
    $M_{\rm 200}$ & $4.14\times 10^{14}$ h$^{-1}$ M$_{\odot}$\\
    $R_{\rm 500}$ & $763$ h$^{-1}$ kpc\\
    $R_{\rm 200}$ & $1178$ h$^{-1}$ kpc\\
    $M_{\rm bh}^{\rm central}$ & $2.17\times 10^{10}$ h$^{-1}$ M$_{\odot}$\\
    \hline
  \end{tabular}
  \caption{Summary of the physical cluster properties used as ICs. These properties were calculated by the on-the-fly FoF and {\sc subfind} algorithms utilised in the original zoom-in simulation.}
  \label{tab:cluster_properties}
  \end{table}
Key physical properties of the cluster calculated by the on-the-fly friends of friends (FoF) and {\sc subfind} algorithms \citep{DavisEtAl85,SpringelEtAl01,DolagEtAl09} at $z\simeq 0.1$ are provided in Table~\ref{tab:cluster_properties}. Our cluster IC is visualised in the central panel of Fig.~\ref{fig:overview}, which shows the temperature encoded by colour and column density encoded by brightness projected through a $15$~h$^{-1}$~Mpc on a side box centred on the central BH of the cluster of interest. Additionally, the radii $r_{500,\:c}=763$~h$^{-1}$~kpc and $r_{200,\:c}=1178$ h$^{-1}$~kpc are indicated by the dot-dashed and dashed lines, each of which contain total masses of $2.82\times 10^{14}$~h$^{-1}$~M$_{\odot}$ and $4.14\times 10^{14}$~h$^{-1}$~M$_{\odot}$, respectively. The figure shows the main cluster as the large region of hot ICM sitting at the nodes of a major cold filament linking to a smaller cluster in the top left of the image and as well as many smaller systems along filaments in other directions that contain many galaxies.

The cluster and epoch was chosen as it shows no signs of a recent AGN outburst and hence provides a relatively ``clean'' environment into which we can inject the jets. Multiple gas phases exist in the simulations including ISM, warm, ICM and jet lobe material. The top- and middle-left panels of Fig.~\ref{fig:overview} show temperature and density projections of a $\sim4\times 10^{10}$ M$_{\odot}$ cold gas structure, face-on and edge-on, respectively. Its disc-like structure presents distortions that are particularly evident in the density projection and has similar properties to the cold gas observed as optical emission line nebula in the centres of some galaxy clusters \citep[e.g.][]{HamerEtAl14,HamerEtAl16,HamerEtAl18}. The structure rotates about the potential minimum of the cluster, as illustrated by line-of-sight velocity contours shown in red (away) and blue (towards) in the middle-left hand panel, and the streamlines of the in-plane velocity shown in the top-left hand panel. 

The chosen jet direction along the $z$-axis is not perfectly perpendicular to the plane of the cold disc, although this outcome is exacerbated by the distorted disc shape. The jet propagation and lobe properties are subsequently altered by jet-ISM interaction \citep{BourneEtAl19}. Additional cold material exists beyond the central cold disc, which contributes to the population of structures that pass near the brightest cluster galaxy (BCG). The bottom-left panel of Fig.~\ref{fig:overview} shows a temperature projection of an in-falling system comprising of multiple cold clumps enshrouded by a bow shock of hot gas. The system initially resides $\sim 65$~h$^{-1}$~kpc from the cluster centre and interacts strongly with the bottom jet lobe at late times.

\section{Results}
\label{sec:results}
\subsection{Lobe evolution}
\label{sec:lobe_evo}
\subsubsection{Overview}
\label{sec:lobe_evo_overview}

\begin{figure*}
\psfig{file=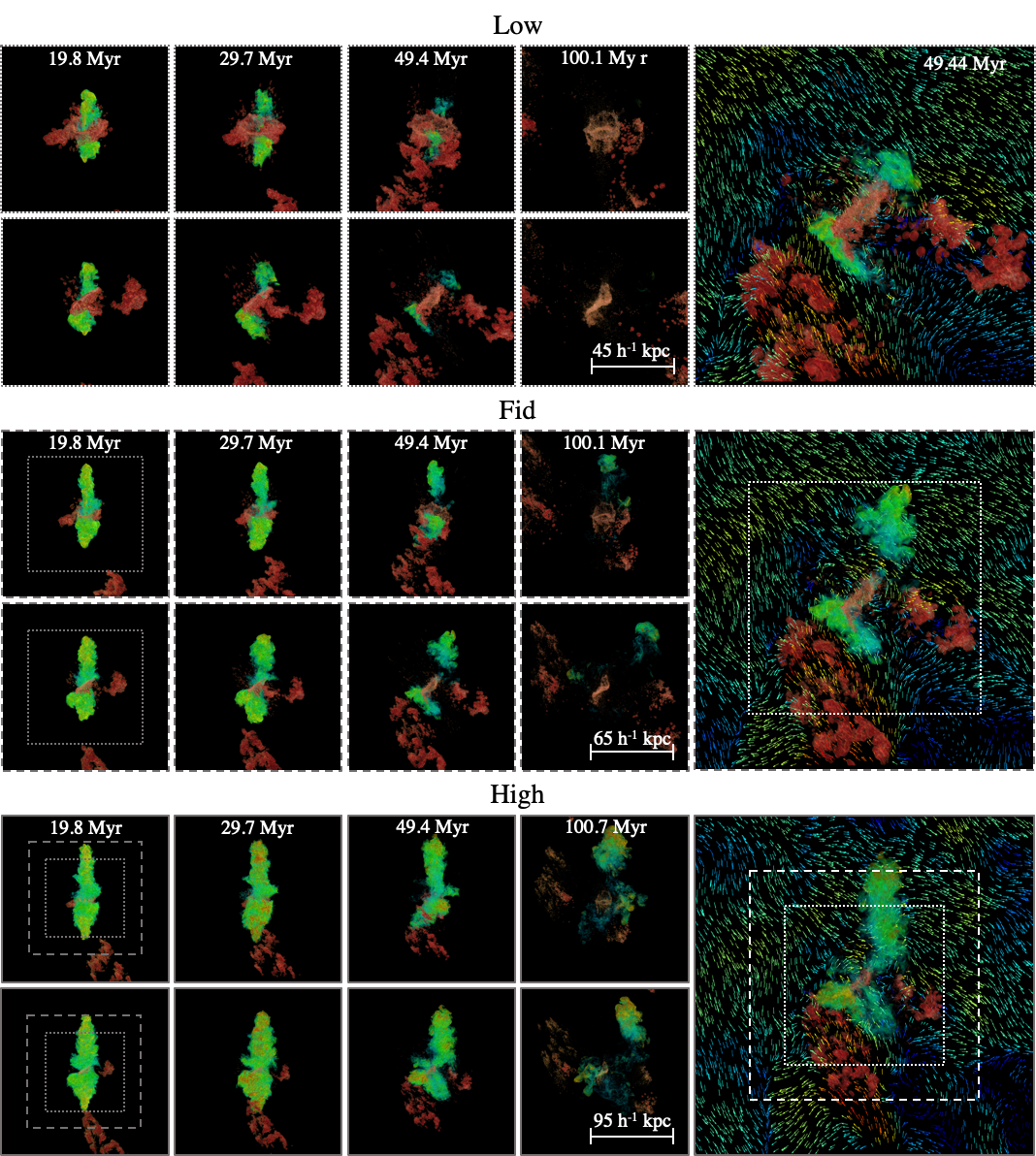,width=1\textwidth,angle=0}
\caption{Volume rendered images of the lobes as illustrated by the jet tracer in green, as well as ISM and warm phase material shown in red. The images are organised into three main groups for each jet power, going from low-to-high power from top to bottom, respectively. Each group consists of eight small panels, which show four different times (left to right) for two line-of-sight orientations (rotated by $90^{\circ}$ about the $z-$axis with respect to each other), and a large panel to the right of each group which shows an enlarged version of the panel at $49.4$~Myr in the lower row, with velocity field lines overlaid. The figure illustrates how cold material interacts with the jet material, in particular the southern lobe, which is disrupted by the passage of a cold structure. The figure also shows the relative survivability of the lobes of different powered jets, with higher power lobes being more resilient to the cluster weather and surviving longer.}
\label{fig:lobe_cold_volume_render}
\end{figure*}

High jet velocities (peaking at $0.56$~c, see Table \ref{tab:jet_runs}), result in strong shocks and lobe material with temperatures of up to $T\sim 10^{11}$~K and densities that can be $\sim 1000$ times lower than the ICM density. This gives rise to observable signatures similar to real clusters such as X-ray cavities surrounded by X-ray bright rims and radio emission from jet material. Mock X-ray and radio observations for different jet powers at $\sim 20$~Myr are shown in the right-hand panels of Fig.~\ref{fig:overview}. X-ray data is produced using {\sc pyXSIM} \citep{ZuHoneEtAl2013} assuming a fixed metallicity of $0.3$ solar and that the cluster is at the same redshift as the Perseus cluster \citep{FabianEtAl06}. Given that we use an effective equation of state to model the ISM and do not include molecular cooling, we are unable to reliably capture the thermal properties of cold dense gas. Therefore, along with setting an upper-temperature limit of $64$~keV, we apply the temperature-density cut of \citet{RasiaEtAl12} to remove gas satisfying $T_{\rm keV} < 3\times~10^6\rho^{0.25}_{\rm cgs}$. RGB composite images are shown in the second column from the right, with each colour representing a different X-ray energy band ($0.5-1.2$, $1.2-2$ and $2-7$~keV, in red, green and blue, respectively). Cooler ICM gas exists in the vicinity of the central disc structure and the previously mentioned in-falling substructure, clearly seen in the low energy band. Cavities surrounded by X-ray bright rims are seen in all three cases but are most obvious in the high power jets, which show large, elongated lobes and clear rims around both the top and bottom lobe. On the other hand, the low power jets produce much smaller and rounder lobes, albeit still with clearly visible rims. Interestingly, while still showing rims around both cavities, for the fiducial and high power jets the top lobe rim is thinner and fainter than for the low power jets. 

The right-hand column zooms in to the cocoon region (indicated by the dashed lines) and shows the high energy X-ray band (blue) overlaid by estimated synchrotron emissivity of jet material (pink). We calculate expected synchrotron emissivity\footnote{Here we are simply interested in morphological features and the relative brightness of regions within each individual object and as such, for simplicity, we do not carry out the full calculation of the radio-spectrum, including free-free absorption, as presented in \citet{BicknellEtAl2018}, which is beyond the scope of this paper.} following Appendix B of \citet{BicknellEtAl2018}, assuming a random magnetic field and that magnetic and relativistic electron energy densities are fixed fractions of the gas internal energy density. There are clear radio bright jet structures along with more diffuse material filling jet lobes. The low power jets have a more distorted morphology and appear bright throughout. The fiducial power jets appear to contain several hot spots, while the high power jets exhibit clear hot spots at their ends. While the high power jets appear FR-II like \citep{FanaroffRiley74}, the low power jets are similar to typical FR-I sources and the fiducial power jets sit somewhat in between, with its extended top lobe but rounder bottom lobe. \citet{EhlertEtal18}, who performed jet simulations in an idealised environment found a transition at a similar power to us, although the high-resolution study of \citet{MassagliaEtAl16} found a transition at a lower power of $P_{\rm J}\sim 10^{43}$~erg~s$^{-1}$. While \citet{EhlertEtal18} found that they could achieve a transition at lower power with improved resolution, we note that at least some of the morphological features observed in our simulations are driven by the interaction of the jet with the cluster environment, most clearly illustrated by the fiducial power run in which the interaction of the bottom jet lobe with the cold disc drives its morphological difference to the top lobe. The radio images also show fluctuations along the jet axis (especially clear in the high and fiducial power jets), consistent with the internal shocks that occur along the jet axis (for further details see Section~\ref{sec:shocks}).

\subsubsection{Lobe dynamics}
\label{sec:lobe_dynamics}

The long term evolution of the jets is shown for all three jet powers in Fig.~\ref{fig:lobe_cold_volume_render}, which comprises volume-rendered images of the jet lobes ($f_{\rm J}>f_{\rm J}^{\rm thresh}$ and $n<0.26$ h$^2$ cm$^{-3}$) as illustrated by the jet tracer in green, while red colours indicate warm and cold ICM material. Each horizontal collection represents a different jet power and contains eight small panels that, from left to right, show the evolution of the system, with the top and bottom rows showing views that are rotated by $90^{\circ}$ about the $z-$axis with respect to each other. The larger panel in each collection shows a clearer view of the jet lobes at $49.4$~Myr along with the in-plane velocity field shown by the coloured arrows. At early times we see the classical double lobe structure emerging above and below the central galaxy, with injected jet material residing largely along the jet axis. By the time the jet switches off there are clear asymmetries between the bottom and top lobes for all three jet powers, with the top lobes typically being elongated and rather narrow, while the bottom jet lobes are shorter and more round, somewhat similar to Abell 4059 \citep{HeinzEtAl02} and Abell 2052 \citep{BlantonEtAl01}. These differences are driven by the interaction of the bottom jet lobes with the cold gas \citep{BourneEtAl19}.

\begin{figure}
\psfig{file=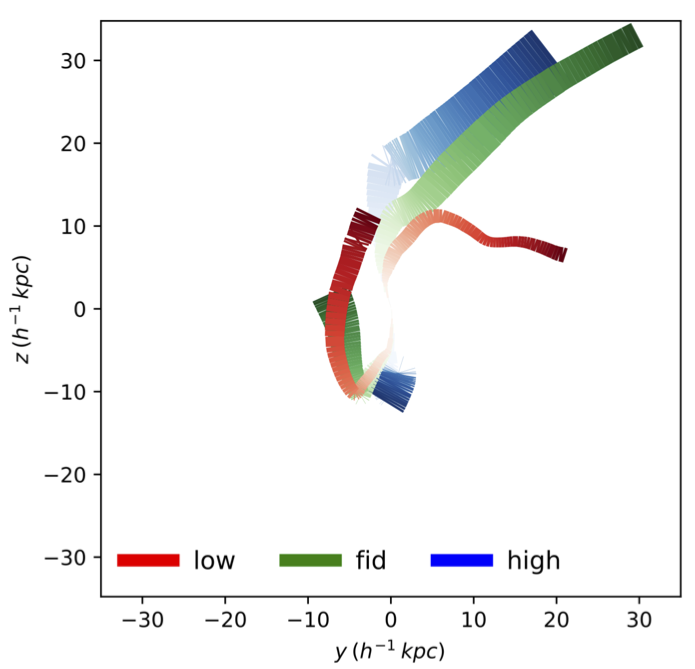,width=0.5\textwidth}
\caption{Overview of the dynamical evolution of the jet lobes for different jet powers, showing the mass-weighted mean position of each jet lobe for the low (red), fiducial (green) and high (blue) power jet runs. Points are calculated for every snapshot and become darker with time, which, on average, are at $1.3$~Myr intervals and in total span a $100$~Myr period. The width of each point shows the dispersion about the mean lobe cell positions, with thicker lines indicating a more extended lobe.}
\label{fig:lobe_dynamics}
\end{figure}

 An alternative comparison of the lobe dynamics for different jet powers is shown in Fig.~\ref{fig:lobe_dynamics}. The plot shows the evolution of the mass-weighted average position of lobe material for the top and bottom lobes in the low (red), fiducial (green) and high (blue) power jet simulations, with averages calculated using the total cell masses over all cells with $f_{\rm J}>f_{\rm J}^{\rm thresh}$ for the relevant jet power. Starting at the origin the lines get darker with time up to a maximum value of $100$~Myr, with each segment representing the value from a single snapshot and the segment width being the mass-weighted standard deviation of the $z-y$ positions of cells within the jet lobes. All three jet powers show somewhat similar behaviour as the lobe material initially travels almost directly vertically along the trajectory of the jets, with the distance travelled increasing with jet power. Once the jet turns off, the top lobe material is displaced to the right by the ram pressure of bulk ICM motions, which is illustrated by the velocity streamlines in the large panels on the right-hand side of Fig.~\ref{fig:lobe_cold_volume_render}. For the fiducial and high power jets this displacement accompanies the continued buoyant rise of the top lobes. However, for the low power jet, the mean $z$ position of the top lobe begins to decrease. Comparing vertical and horizontal motions, we find that the horizontal motion becomes increasingly more important compared to the buoyant motion of the top lobe for lower power jets. After $\sim 40-50$~Myr, a shredded cold substructure, directly interacts with the bottom lobe, first compressing and then disrupting it. We find that the average trajectory of the bottom lobe material in the low and fiducial power jets is completely reversed, while for the high power jet its progress appears to stall. The jet lobe sizes and lifetimes increase with jet power: while the low power jet lobes are completely absent from the $100$~Myr panel, and very little of the fiducial power jet lobes remain (mostly the top lobe), having been completely destroyed by the ICM weather and mixed in with the ambient medium, the top lobe of the high power jet run still appears prominently along with the diffuse remains of the bottom lobe. Overall, these two figures illustrate the significant role that the inherent ICM velocity field can have on the trajectories of jet lobes. Such displacement of lobes due to ICM bulk motions have been seen in previous works \citep{BourneSijacki17, SijackiEtAl09} and could provide an explanation as to why relic lobes appear along different trajectories without invoking large-angle jet precession \citep[see e.g.][]{BabulEtAl13,DunnEtAl06}.

\subsubsection{Lobe energetics}
\label{sec:lobe_energetics}
\begin{figure*}
\psfig{file=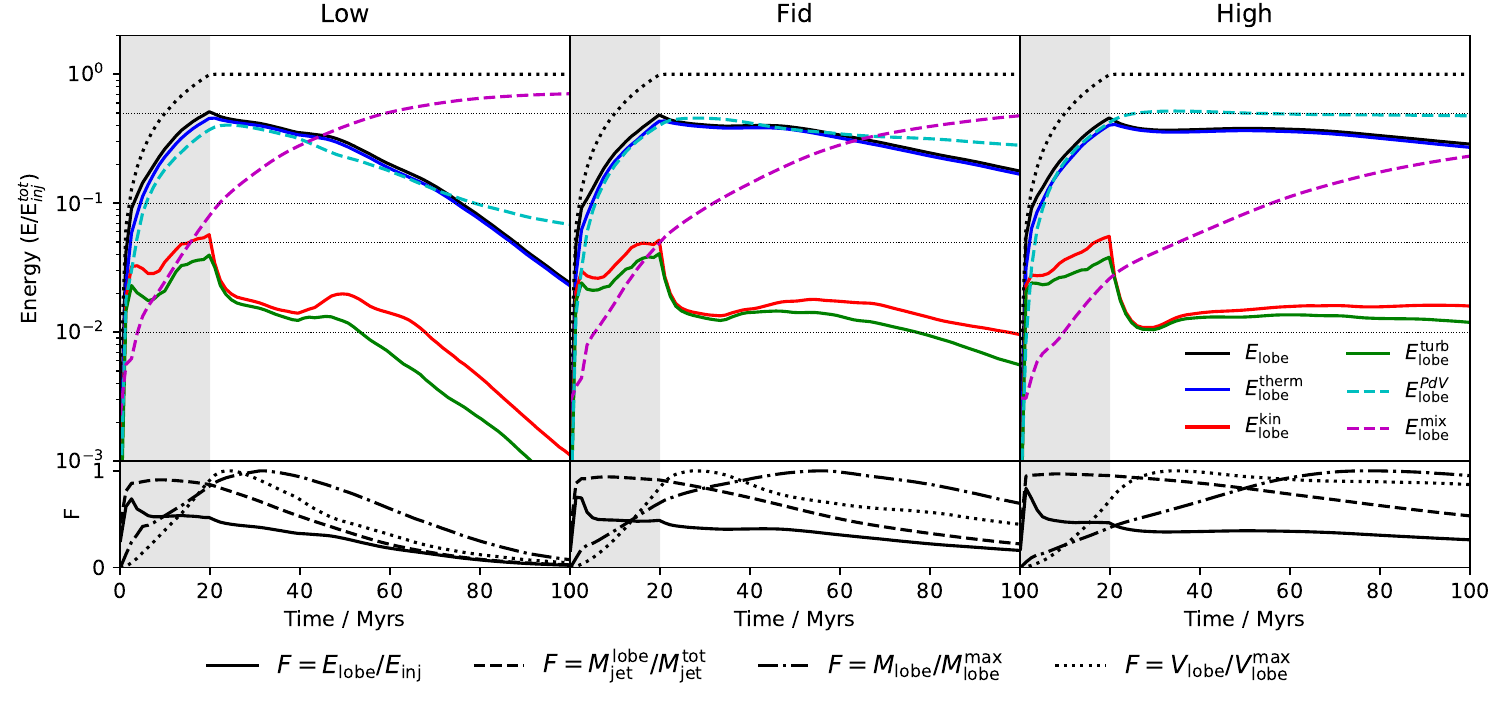,width=1\textwidth,angle=0}
\caption{Evolution of jet lobe energy content is shown in the {\it top panel}, split into the thermal (blue), total kinetic (red) and turbulent (green) components. The total injected energy is shown by the dotted line, while the total (thermal+kinetic) energy content of jet lobe material ($f_{\rm J}>f_{\rm J}^{\rm thresh}$) is shown by the solid black line. The dashed cyan and magenta lines show estimated lobe energy loss due to $PdV$ work and mixing, respectively. While the jet is active $\sim$half of the total injected energy remains in the lobe and $\sim$ half goes into the ICM. The grey shaded region indicates the period over which the jet is active. The thermal component dominates the lobe energy, while the turbulent component dominates the total kinetic energy. In the {\it bottom panel}, we show the evolution of four mixing related quantities, namely the lobe energy normalised by the total injected energy (solid), the injected jet mass within the lobe normalised by the total injected mass (dashed), the total lobe mass normalised to its maximum value (dot-dashed), and the lobe volume normalised by its maximum value (dotted).}
\label{fig:lobe_energy_evo}
\end{figure*}

Tracking the energy content of lobes and the $PdV$ work done to inflate them can provide insight into ICM heating mechanisms and allow comparison with observational studies that estimate jet energetics from lobe properties \citep[e.g.][]{RaffertyEtAl06,McNamara2007,NulsenEtAl07,DunnFabian08,Fabian12}. The main panels in Fig.~\ref{fig:lobe_energy_evo} show lobe energy evolution for each jet power, normalised by the total injected energy ($\int_{0}^{20\:{\rm Myr}}P_{\rm J}dt$). The combined lobe thermal plus kinetic energy is shown by the solid black line, the individual thermal, total kinetic and turbulent kinetic energies are shown by the blue, red and green solid lines, respectively, and the total injected energy by the dotted black line. Estimated lobe losses due to $PdV$ work and mixing are shown by the dashed cyan and magenta lines, respectively. Quantities are calculated following the method in \cite{BourneEtAl19} assuming jet bulk velocity cut-offs given by $v_{\rm J}^{\rm mean}$ defined in Table~\ref{tab:jet_runs}. The grey shaded region indicates the period over which the jet is active and the horizontal dotted lines indicate $1$, $5$, $10$ and $50\%$ of the total injected energy. Additionally, the three lower panels show the evolution of other lobe properties of interest, namely the ratios of the lobe energy to injected energy ($E_{\rm lobe}/E_{\rm inj}$, solid line), tracer mass in the lobe to total tracer mass ($M_{\rm J}^{\rm lobe}/M_{\rm J}^{\rm tot}$, dashed line), total lobe mass to maximum lobe mass ($M_{\rm lobe}/M_{\rm lobe}^{\rm max}$, dot-dashed line) and lobe volume to maximum lobe volume ($V_{\rm lobe}/V_{\rm lobe}^{\rm max}$, dotted line).

During the inflation phase, we find that the combined thermal and kinetic energy of the lobes generally accounts for $\sim$~half of the injected jet energy \citep[see also e.g.,][]{HardcastleKrause13, BourneSijacki17, WeinbergerEtAl17}, with values of $51.7$, $48.9$ and $46.3\%$ for the low, fiducial and high power jet lobes, respectively, shortly before the jets switch off. Large departure from this only occur in the first few Myr of jet injection, where the fraction of injected energy in the lobes peaks at $70-80\%$ before significant expansion and mixing has occurred. For all three jet powers, the thermal component dominates the energy budget because the shocks driven by the fast kinetically dominated jets rapidly heating the lobe gas. The residual kinetic energy accounts for only $\sim 5-6\%$ of the jet energy residing in the lobes by $20$~Myr, which itself is dominated by the turbulent component that accounts for $\sim 70-80\%$ of the lobes total kinetic energy.  

Because lobes retain only half of the jet energy during inflation, the rest must be communicated to the ICM. Based on our simulations, we find that mixing is sub-dominant during the inflation phase, contributing between $8\%$ of the injected energy from the low power jet down to $2.6\%$ for the high power jet. On the other hand, the $PdV$ work done to inflate the lobes accounts for between $37.4\%$ and $42.5\%$ of the injected energy from low to high jet powers, respectively. The amount of $PdV$ work done on the ICM and how this energy manifests itself within the ICM depends on how rapid or ``explosive'' the lobe inflation is (see Section~$5.4$ of \citet{McNamara2007}). If inflation is slow (adiabatic), the ICM can be compressed and sound waves can be driven. If inflation is fast, shocks are driven into the ICM, generating entropy as kinetic energy is irreversibly dissipated as heat. Fig.~\ref{fig:lobe_energy_evo} shows that the fraction of injected energy going into $PdV$ work increases with jet power, inline with theoretical expectations. 

At the end of inflation, the ratio of the instantaneous $PV$ estimate, i.e. using the instantaneous lobe pressure and volume, and integrated $PdV$ estimate are $PV:\int PdV=0.82$, $0.72$ and $0.64$ for the low, fiducial and high jets powers, respectively. Therefore, $PV\sim\int PdV$ only holds if the lobe pressure remains roughly in equilibrium with the ICM pressure, i.e. for slow adiabatic inflation of the lobes. This assumption breaks down at higher powers and impacts our ability to estimate jet powers and the energies needed to inflate lobes from observations. The instantaneous lobe enthalpy, $H=E_{\rm lobe}^{\rm therm}+PV=\gamma PV/(\gamma - 1)$, is often taken as a measure of the jet energy needed to inflate a lobe. We find that the ratio of $H:E_{\rm J}^{\rm inj}=0.77$, $0.73$ and $0.68$ for the low, fiducial and high power jets, respectively. I.e., the lobe enthalpy underestimates the energy needed to inflate the lobes by at least $23\%$ in the low power case and $32\%$ for the high power jets, suggesting that lobe enthalpy sets a lower limit on the required energy. 

While the partition of energy for different jet powers are quite similar during lobe inflation, differences become more apparent after jets switch off. All runs show a sudden drop in the kinetic energy component as the jet action stops, accompanied by a drop in the thermal energy component, which is no longer replenished by shocked jet material. Significant differences in the fraction of energy lost due to mixing begin to appear between the different jet powers, with the low power lobes being more susceptible to mixing than the higher power lobes. Although it has been suggested that jet-driven instabilities can effectively mix lobe material with the ICM \citep{HillelSoker16, HillelSoker17}, we previously showed that by implementing appropriate refinement criteria our method results in limited instability and numerical mixing \citep{BourneSijacki17}. We therefore stress that the efficient mixing seen at late times is predominantly driven by ICM and substructure motions disrupting the lobes. By $100$~Myr we estimate that while only $23.4\%$ of the high power jets energy has been mixed into the ICM, this jumps up to $71\%$ in the case of the low power jets, suggesting that the lower power jet is more susceptible to the cluster weather. While different values of $f_{\rm J}^{\rm thresh}$ chosen for each jet power can impact the mixing rate, a more detailed mixing analysis in Section~\ref{sec:mixing}, shows that our conclusion is still valid. The stark differences in the levels of mixing lead to stark differences in the amount of energy retained by the jet lobes, with the low power jet lobes retaining only $2.4\%$ of the injected energy by $100$~Myr, while the high power jet retains $28.6\%$. 

Finally, while the cumulative $PdV$ remains flat after the high power jets switch off, it decreases for lower power jets, suggesting that work is done on the jet lobes. From the lower panels, peak lobe volumes occur between a few to $\sim 15$~Myr after the jet action halts, while peak lobe masses occur later. These peaks occur earlier for lower power jets and the subsequent declines appear steeper when compared to the higher power jets. Continued entrainment and hence gain in mass even after lobe volumes begin to decrease suggests that the lobes are compressed, leading to the decrease in cumulative $PdV$ losses. The effect is particularly evident around $\sim 40$~Myr, where lobe thermal energy shows a distinct {\it bump} and corresponding dip in $PdV$ losses. This is due to the in-falling substructure (see Section~\ref{sec:lobe_dynamics}) impacting the bottom lobe. However, compression is not the only mechanism that can effect our $PdV$ estimate. Once lobes begin to lose significant amounts of mass, the corresponding volume of the lost material will also impact our $dV$ estimates and as such, we suggest that ours estimates begin to break down as the system evolves beyond the mass peak. 

\subsection{Energetics: from jet to environment}
\label{sec:icm_heating}
\subsubsection{Where does the energy go?}
\label{sec:icm_heating_energy}
\begin{figure*}
\psfig{file=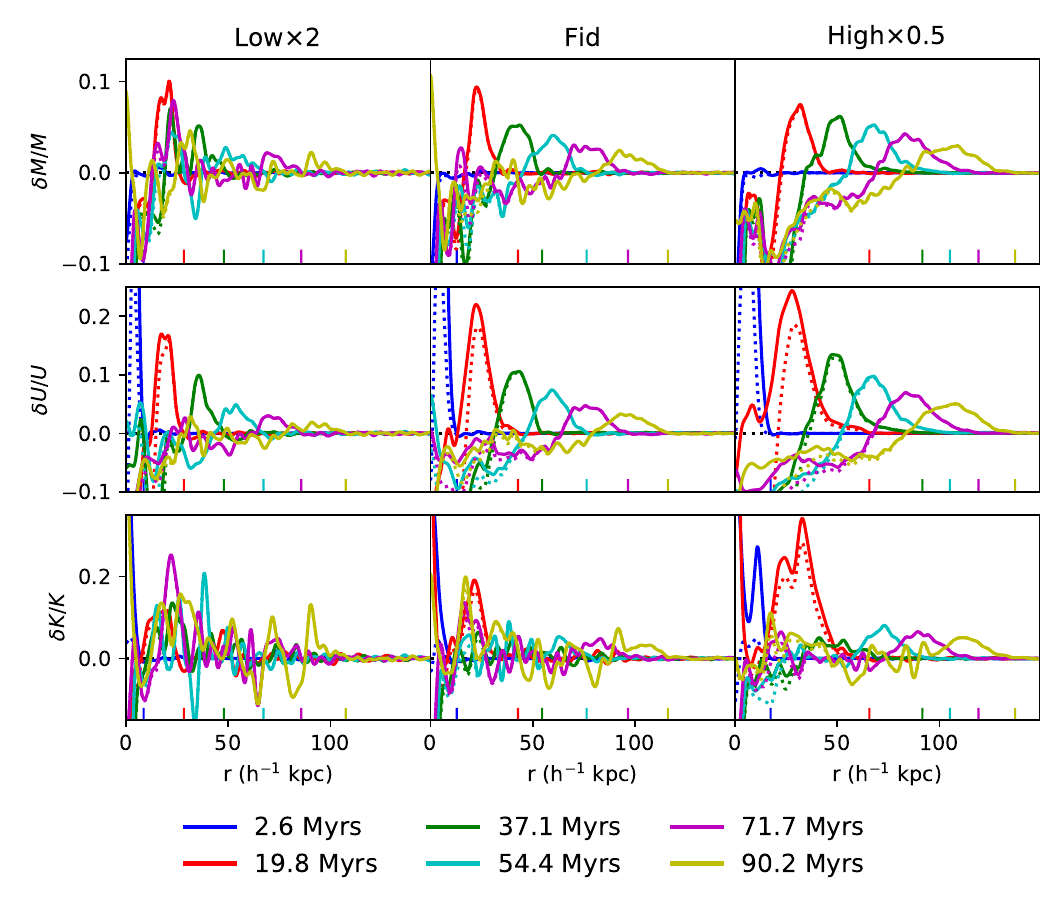,width=1.\textwidth,angle=0}
\caption{Changes in radially averaged quantities between jet and non-jet runs. From top to bottom, the rows show the radial fractional difference in mass, internal energy and kinetic energy, with each jet power, from low to high, shown in each column from left to right, respectively. The profiles are calculated at different times and shown in different colours, as given by the key, and are presented both with and without jet lobe material included, by the solid and dotted lines, respectively. Additionally, the low/high power run quantities have been doubled/halved, to allow all three jet power runs to be plotted on the same $y-$axis and we have defined a time-dependent {\it influence} radius, $r_{\rm out}$, indicated by the small ticks at the base of each panel and defined by the outer edge of the main $\delta U/U$ peak.}
\label{fig:icm_energy_evo}
\end{figure*}

\begin{table}  
\centering
\begin{tabular}{|l|l|l|l|l|l|l|l|}
\hline
Run & $\frac{r_{\rm out}}{{\rm h}^{-1}\:{\rm kpc}}$ & $\frac{\Delta E_{\rm therm}}{E_{\rm inj}^{\rm tot}}$ & $\frac{\Delta E_{\rm kin}}{E_{\rm inj}^{\rm tot}}$ & $\frac{\Delta E_{\rm pot}}{E_{\rm inj}^{\rm tot}}$ \\
\hline
low & $108$ &$6.0\pm 5.4\%$ & $28.1\pm 1.7\%$ & $72.3\pm 22.5\%$ \\
fid & $116$ &$40.8\pm 1.5\%$ & $8.8\pm 0.3\%$ & $55.6\pm 4.5\%$ \\
high & $137$ &$58.5\pm 1.2\%$ & $8.7\pm 0.1\%$ & $52.7\pm 1.4\%$ \\
\hline
\end{tabular}
\caption{Differences in the thermal (gas), kinetic (gas) and potential (gas + stars) energy between runs with and without jets, calculated within $r_{\rm out}$ at $90.2$~Myr and normalised by the total injected jet energy for the corresponding simulation. Note that $r_{\rm out}$ is defined as the outer radius of the main peak in $\delta U/U$ as shown in Fig.~\ref{fig:icm_energy_evo}.}
\label{tab:delta_E}
\end{table}  

 Differences in the energy content of any two systems is a combination of energy directly injected by the jet and differences in energy gains/losses that arise from the other physical processes, such as radiative cooling, star formation, and stellar feedback, that are themselves impacted by the jet action. Therefore, the energy difference between a jet and non-jet run could exceed the total energy input of the jet, as is the case for the high power jet run (see Table~\ref{tab:delta_E}). In addition to the non-linear nature of the physical processes and their inter-dependant evolution, the stochastic nature of simulations and the effect that small perturbations (such launching a small scale jet) can have on the system\footnote{For example, when a jet is launched, the constituent cells require a short time-step. Criteria that limit the relative positions of neighbour cells on the time-step hierarchy can result in the propagation of time-steps far from the jet itself that are smaller than in the run without a jet.} can also result in differences between runs \citep[see e.g.,][]{KellerEtAl2019}. While such differences appear as only very small fluctuations when compared to the whole system, they can be large compared to the total injected jet energy. 

To explore where the jet energy ultimately resides in the system, we compare the evolution of the fractional differences in mass and energy profiles of the clusters with and without jets in Fig.~\ref{fig:icm_energy_evo}. The rows, from top to bottom, show the fractional difference in mass, thermal energy and kinetic energy as a function of radius, and from left to right for the low, fiducial and high power jets, respectively. Values calculated for low/high power jets have been doubled/halved to fit all three jet powers on the same $y$-axis and all curves have been smoothed using a Gaussian filter with a standard deviation of $1.5$~h$^{-1}$~kpc to reduce noise. The different coloured lines refer to different times, while the solid and dotted lines show values calculated with and without jet material included, respectively. All quantities are calculated within radial bins of fixed width $dr=1.5$~h$^{-1}$~kpc and the small vertical lines at the bottom of each panel define an {\it influence} radius, $r_{\rm out}$, calculated as the outer radius after the main peak in $\delta U/U$, i.e. the radius at which it returns to $zero$.

The qualitative behaviour is similar for all fractional changes, albeit with higher power jets exhibiting more well defined, larger (noting the re-normalisation), broader and longer lived changes. The inflation and subsequent rise of the lobes sweeps up and displaces gas to larger radii, as indicated by the positive peak in $\delta M/M$ moving outwards and flattening with time and reduced values at small radii. Similarities in the $\delta M/M$ curves with and without jet-material illustrate that the displaced mass is dominated by non-jet material, with higher power jets displacing more mass. At early times the main peaks exist at $r\simlt r_{\rm lobe}^{\rm max}$, where $r_{\rm lobe}^{\rm max}$ is the maximum radius of lobe material, however, at later times (particularly for low and fiducial power jets) the peaks occur beyond $r_{\rm lobe}^{\rm max}$ and may be related to the weak shock/sound wave generated by the initial lobe inflation. While in some cases, particular for low power jets, small fluctuations are present beyond the main peak, they are very subdominant compared to the apparent influence of the jet. 

Mass displacement by jet feedback leads to a net gain in the potential energy of baryonic material within the central region of the cluster. To quantify and minimise the effect that small fluctuations in $\delta M/M$ seen at large radii may have on the net difference in potential energy of a given jet run compared the non-jet run, $\Delta E_{\rm pot}$, we perform averaging over spheres. Specifically, we calculate $\Delta E_{\rm pot}$ within multiple spheres that have outer radii ranging from $r_{\rm out}$ to $1.25 \times r_{\rm out}$ at $1.5$~h$^{-1}$~kpc intervals (matching the bin widths in Fig.~\ref{fig:icm_energy_evo}), and take the standard deviation in these values as the associated error.

Average values of $r_{\rm out}$ and $\Delta E_{\rm pot}/E_{\rm tot}^{\rm inj}$ at $90.2$~Myr are shown in the second and final columns of Table~\ref{tab:delta_E}, respectively. The associated error is large for the low power jets and decreases for higher powers, although this may be expected given that the values are a percentage of the total injected energy. However, despite the large error for the low power jet, the average value is still consistent with that of the fiducial and high power jets, which implies that systems including jets exhibit potential energies that are the  larger than those of non-jet systems by $\simgt 50\%$ $E_{\rm inj}^{\rm tot}$.

Previous simulations found smaller values, for example in  \citep{BourneSijacki17}, we found for a non-radiative run $\sim 16\%$ of the inject energy ends up as potential energy \citep[see also][]{HardcastleKrause13}, however, these simulations only considered values while the jet is active and did not capture the buoyant rise and further displacement of mass after the jets switch off. On the other hand, \cite{WeinbergerEtAl17} performed simulations that tracked the post-inflation rise of the lobes and found that $\sim 40\%$ of the jet energy goes into the potential energy of the system, much closer to the values found here. While it has been suggested that most of the jet energy ultimately ends up in potential energy of the system \citep[e.g.,][]{Matthews2009, MatthewsGuo2011, McNamara2007}, it is important to remember that we measure the net difference between two systems (jet vs. non-jet) at one instance in time, capturing the combined effect of the lobe inflation displacing mass outwards and the lack of jet heating possibly resulting in mass flowing inwards. Instead, measuring the change in the potential energy within $r<150$~h$^{-1}$~kpc between $t=0$ and $t=90.2$~Myr separately for each run, we find net losses in potential energy irrespective of the inclusion of jets feedback, because ultimately the jets only acts to reduce the net inflow of gas \citep[see also][]{GuoEtAl2018}. 

Considering all gas (solid lines), the similarities between the troughs in $\delta U/U$ and $\delta M/M$ indicate that at small radii the drop in energy is due to the removal of mass. However, larger peaks in $\delta U/U$ compared to those in $\delta M/M$ indicate that the gains in thermal energy here are not caused by mass redistribution alone and must also include heating due to the jet action. When jet material is not included, there are differences between the $\delta U/U$ curves, which are more pronounced for higher power jets at earlier times and at smaller radii, where the energy in the jet material makes a significant contribution to the total energy budget. This fits with the picture drawn in Section~\ref{sec:lobe_energetics}, whereby during inflation, the jet lobes retain a significant fraction of energy within the cluster centre. Thereafter this energy is transferred to the ICM at ever larger radii as the lobes rise and become disrupted by the cluster weather. As for $\delta M/M$, the prominent peaks correspond well to the outer regions of the jet feedback influence and any fluctuations beyond this region are small compared to the total thermal energy of the environment. 

Differences in thermal energy content between jet and non-jet runs, $\Delta E_{\rm therm}/E_{\rm inj}^{\rm tot}$, are estimated in the same manner as the potential energy and are listed in the third column of Table~\ref{tab:delta_E}. The associated error is largest for the low power jet, however, it is considerably smaller than for the potential energy estimate. At face value, we find the difference in thermal energy between jet and non-jet runs, as a percentage of jet power, increases with jet power. In other words, more powerful jets are more efficient at heating the ICM. This makes sense physically for a couple of reasons. First, higher power jets drive stronger shocks into the ICM and hence not only convert more of their energy into thermal energy, but also heat gas to higher temperatures that has longer cooling times and therefore remains in the system longer. Second, short ICM cooling times mean that thermal energy  in lobe material mixed into the ICM would radiate more quickly than if it were retained in the lobes. From Fig.~\ref{fig:lobe_energy_evo}, we found that higher power jet lobes retain a larger fraction of the injected energy at late times and the longer that energy can be retained in the low density, high temperature jet lobes the more thermal energy the system is likely to retain. 

$\delta K/K$ shows far more fluctuations and although for high jet powers the peaks behave similarly to the fractional $\delta M/M$ and $\delta U/U$, with larger peaks forming at small radii early on before moving to larger radii and flattening with time, this behaviour is less obvious at later times in the fiducial power jet run, and is essentially non-existent in the low power jet run. Additionally, fluctuations seen at radii beyond $r_{\rm out}$ sometimes appear rather large, e.g., the red bump in the low power run just below $50$~h$^{-1}$~kpc. This suggests that differences driven by stochasticity of the simulation \citep{KellerEtAl2019} (for example the possibility of substructures on slightly different orbits in different runs) potentially have a greater effect on the kinetic energy compared to other quantities.

With this caveat in mind we present $\Delta E_{\rm kin}/E_{\rm inj}^{\rm tot}$ at $90.2$~Myr in Table~\ref{tab:delta_E}, but caution that although by this time fluctuations beyond $r_{\rm out}$ appear small, larger early time fluctuations may now reside within $r_{\rm out}$ and contribute to our estimates of $\Delta E_{\rm kin}$. This could explain why the low power jet run seemingly contains a larger kinetic fraction than the fiducial and high power jet runs, which both contain $\sim 8-9\%$ of $E_{\rm inj}^{\rm tot}$. On the other hand, if these values are solely driven by jet feedback, a possible physical explanation is that higher power jets result in lower kinetic energy differences (and higher thermal energy differences) because they thermalise a larger fraction of their kinetic energy through shocks, and are more effective at inhibiting/thermalising bulk flows within the ICM (for example, see right hand panels of Fig.~\ref{fig:lobe_cold_volume_render}, where the ICM velocity fields are more readily disrupted by higher power jet lobes). However, to definitively confirm whether or not this picture is correct, it will be necessary to perform simulations covering a wider range in parameter space in terms of both jet power and cluster environment, which is beyond the scope of present work. 

\subsubsection{Shocks and sound waves}
\label{sec:shocks_and_sounds}
\label{sec:shocks}
\begin{figure*}
\psfig{file=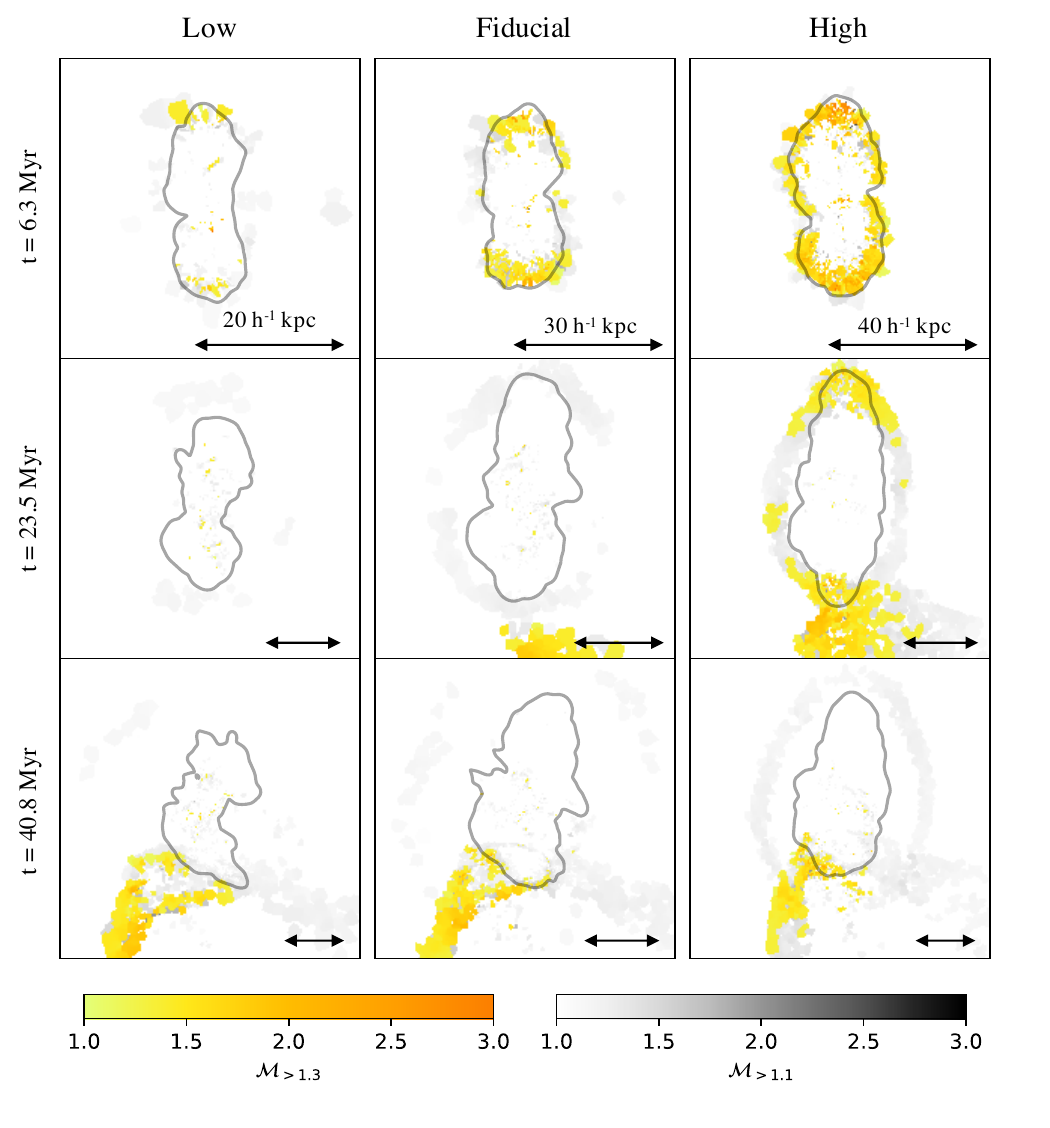,width=1\textwidth,angle=0}
\caption{Dissipation-weighted Mach number projections, showing different jet powers, low to high, in each column from left to right, respectively, and at different times, increasing down each column. The projections are performed through boxes with a depth equal to $1/5$th the box width, for two different shock finder setups that correspond to the {\it standard} finder in yellow/orange and {\it weak shock} finder in grey-scale. The figure illustrates that higher power jets typically produce stronger shocks for longer. It is also worth noting that in all runs the weak shock finder detects a feature propagating away from the jet lobes that is consistent with a sound wave driven into the ICM.}
\label{fig:mach_projection}
\end{figure*}

In this section, we look in detail at the work done by inflating lobes against the ICM, discuss where and when shocks and sound waves are important, and how they compare for differ power jets. We use the shock finder of \cite{SchaalSpringel15}, where cells need to meet three key conditions to constitute the shock zone. Namely, the finder must detect compression ($\nabla\cdot{\bf v}<0$), guard against misclassifying weak shocks and contact discontinuities (requiring $\Delta\log T\geqslant\log [T_2/T_1]|_{\mathcal{M}=\mathcal{M}_{\rm min}}$ and $\Delta\log P\geqslant\log[ P_2/P_1]|_{\mathcal{M}=\mathcal{M}_{\rm min}}$), and avoid detecting spurious shocks such as in shear flows (requiring $\nabla T\cdot\nabla\rho>0$). 

Fig.~\ref{fig:mach_projection} shows dissipation-weighted Mach number projections for low-to-high jet powers and at different times. Projections are performed through a distance of one-fifth of the box width. \cite{SchaalSpringel15} set a minimum Mach number of $\mathcal{M}_{\rm min}=1.3$ when applying the criteria outlined above and our results using this setting are shown in the yellow-orange colour maps. Additionally, we attempt to find weaker shocks using $\mathcal{M}_{\rm min}=1.1$, as shown by the grey-scale colour maps. The footprint of the jet lobe material is shown by the solid grey contour. At early times the jets undergo internal shocks and drive bow shocks, inflating lobes of hot plasma that continue to expand into the ICM. The initial rapid expansion of the lobes slows and shocks driven into the ICM weaken and broaden into sounds waves. At late times, strong shocks are again visible at the base of the bottom lobes as the substructure mentioned previously moves through the ICM and interacts with the bottom lobe. 

As jet power increases the quantity and strength of shocks increases, with the low power jets showing very few strong shocks at early times only, and high power jets showing strong shocks throughout the jets' lifetime and even for a short period after the jets switch off. All jet powers show weak shocks, which at later times {\it detach} from the lobes and travel at roughly the local sound speed in all directions, suggesting that the shocks have broadened into sound waves. The Mach numbers of shocks driven into the ICM in our simulations are consistent with those observed in real galaxy clusters \citep[e.g.,][]{FabianEtAl06, FormanEtAl07, CrostonEtAl11, SandersEtAl2016, GrahamEtAl08} and the behaviour found between jet power and shock strength is in agreement with other simulations \citep[e.g.,][]{EhlertEtal18}.

\begin{figure*}
\psfig{file=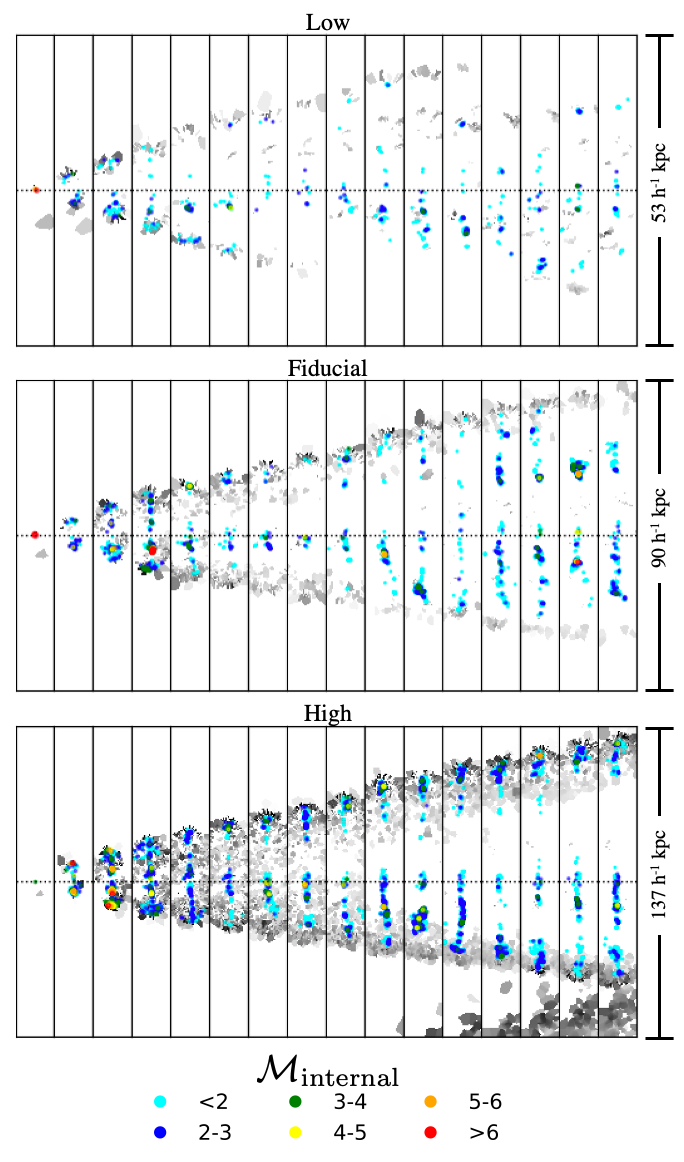,width=0.72\textwidth,angle=0}
\caption{Detection of internal shocks within the jet lobe material. Grey-scale maps show emission-weighted Mach number projections (using $\mathcal{M}_{\rm min}=1.3$), while the coloured points show the locations of shocks detected within jet lobe material. Each row shows a different jet power run while each column shows a different time spanning from $t=0$ to $19.8$~Myr at $\sim 1.3$~Myr intervals. Additionally, the dotted line shows the horizontal intersection of the central black hole. Higher power jets drive more internal shocks that typically have higher Mach numbers.}
\label{fig:internal_shock_points}
\end{figure*}

Shocks can play two distinct roles, namely {\it i)} thermalising the kinetic jet and inflating the lobes, and {\it ii)} heating the ICM due to shocks driven by the lobe expansion. To understand the first of these processes, we consider shocks that occur within jet material (based on the jet tracer definition), which we call ``internal'' shocks, versus those that occur in non-jet material. In Fig.~\ref{fig:internal_shock_points}, we plot dissipation-weighted Mach number projections in grey for the period over which the jet is active. Additionally, we add points to indicate the locations of internal shocks colour coded by their Mach number. This is done for all three jet powers using the results from the $\mathcal{M}_{\rm min}=1.3$ shock finder, and are shown separately in each row as labelled.

Strong internal shocks typically occur at earlier times, with only a handful having $\mathcal{M}>6$ and the majority of shocks are not particularly strong, i.e. many have $\mathcal{M}<2$ and most have $\mathcal{M}<3$. However, the fact that the shocks have relatively low Mach numbers, despite the very high velocities observed in the jets is because of the high sound speeds within the hot lobes, which in itself is a result of the earlier strong shock heating, meaning that Mach numbers are typically low.

More shocks occur as the jet power is increased. For the low power jets, it seems that internal shocks appear somewhat sporadically and do not extend along the entire jet (in some cases ending well before the bow shock). However, for the fiducial and high power jets the internal shocks are typically seen along the whole jet, and particularly in the high power jets reach the region of the bow-shocks (although appear confined to the jet axis). Note further, that while a large number of shocked cells in the bow shocks are not classified as being within the jet material, they could still be important in providing thermal energy to the jet lobes via mixing (see below).

\begin{figure*}
\psfig{file=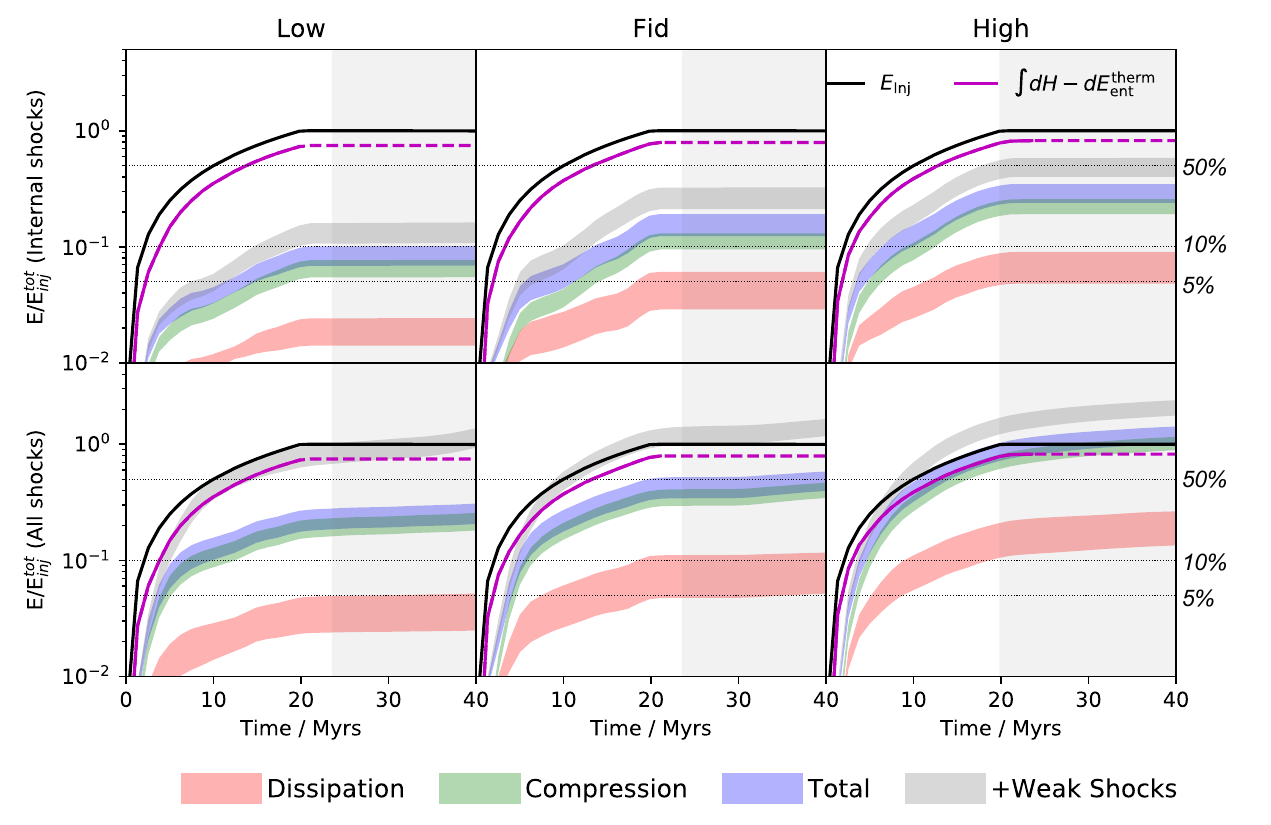,width=1\textwidth,angle=0}
\caption{Evolution of the shock energy budget compared to the jet energy budget. The top row shows the evolution for internal shocks only, while the bottom row shows the evolution for all shocks in the vicinity of the jet and lobes (i.e. $f_{\rm J}>10^{-15}$). In each case, the energy budget has been split up into the dissipation (red), adiabatic compression (green) and total (blue). An additional band (light grey) shows total energy fractions if we include weak shocks, as calculated by using the shock finder with $\mathcal{M}_{\rm min} = 1.1$. The bands are calculated by detecting shocks via pressure (lower bound) or temperature (upper bound) jumps. All panels show the cumulative injected jet energy (black line) and the minimum required energy to inflate the jet lobes (magenta line). Internal shocks make a greater contribution to the total lobe energy as jets become more powerful, but even the high power jet requires gas that has been shocked in the bow shock to contribute to the lobe thermal energy. Additionally, the fraction of shock heating (and the strength of the shocks involved) available to heat the ICM increases with jet power.}
\label{fig:shock_energy_evo}
\end{figure*}

Fig.~\ref{fig:shock_energy_evo} shows how the total shock energy produced by the jets evolves. The low, fiducial and high power jet runs are shown from left to right, respectively, while the top and bottom rows show ``internal'' shocks and all shocks, respectively. \cite{SchaalSpringel15}, estimate shock energy using Mach numbers defined through the shock temperature jumps. However, due to the possibility of confusing contact discontinuities or weak shocks with strong shocks between the hot lobes and the shocked ambient medium, which exhibit large temperature jumps, we additionally calculate shock energies using pressure jumps. This approach generally results in lower Mach number estimates and hence lower shock energies. Therefore, in Fig.~\ref{fig:shock_energy_evo} we present shock energies as coloured bands that span between these two sets of estimates. We separate out dissipative heating, {which results in the irreversible conversion of kinetic energy to heat} and adiabatic compression, as shown by the red and green bands, respectively. The blue band shows the combination of the two, calculated using the minimum Mach value of $\mathcal{M}_{\rm min}=1.3$, which due to the reversible nature of adiabatic processes does not reflect the total thermal energy gained by the system, but rather the total total thermal energy produced via shocks over the lifetime of the simulation. The dark grey band additionally shows the total shock thermal energy budget if we instead assume $\mathcal{M}_{\rm min}=1.1$. Finally, we only include cells with $f_{\rm J}>10^{-15}$. Having tested a range of values we find that this choice suitably limits the detection of shocks driven by processes at large distances and removes the late time sound wave highlighted in Fig.~\ref{fig:mach_projection}, without compromising the detection of shocks close to the jet lobes.

Jets initially comprise almost exclusively of kinetic energy, however, from Fig.~\ref{fig:lobe_energy_evo}, the vast majority of this energy must rapidly thermalise. The minimum amount of thermalisation required is equal to the thermal energy in the lobes minus that of entrained material, plus the energy needed to inflate the lobes. We estimate this as $E_{\rm lobe}^{\rm req}=\int dH - dE_{\rm ent}^{therm}$, where $dH=dE_{\rm lobe}^{\rm therm}+PdV$ is the differential lobe enthalpy, and $dE_{\rm ent}^{therm}=\bar{u}_{\rm ICM} dm_{\rm ent}$ is the estimated thermal energy of the entrained ambient material. This estimate is calculated until shortly after $20$~Myr, at which point the jets stop and $E_{\rm lobe}^{\rm req}$ reaches its maximum value, and, along with the cumulative injected jet energy is shown in Fig.~\ref{fig:shock_energy_evo} by the magenta and black lines, respectively. As in Fig.~\ref{fig:lobe_energy_evo} all quantities are normalised by the total injected energy for easy comparison between different jet power runs. 

The top row shows the total thermal energy injection by {\it internal} shocks, we see that while quantitatively they differ, the general behaviour is similar for all jet powers; initially, we measure limited thermal energy injection via shocks, before a rapid increase after a $\sim$few~Myr, followed by a more steady increase in the total energy injection, which eventually plateaus as the jet action halts and no further {\it internal} shocks are detected. Additionally, we find that the total shock thermal energy (blue band), is dominated by adiabatic compression as opposed to direct dissipation of kinetic to thermal energy in the shock, which, is a result of the majority of the shocks having Mach numbers below $3$ (roughly the point at which the contribution due to adiabatic compression is equal to that due to dissipation). 

None of the jet powers provides sufficient thermal energy via {\it internal} shocks to meet the required conversion of jet kinetic energy into lobe thermal energy and $PdV$ work. If we include weak shocks, the high power jets come closest, with {\it internal} shocks accounting for at most $\sim 74\%$ of the required energy, while weak power jets only account for up to $\sim 25\%$ of the required energy through {\it internal} shocks. This suggests that shocks detected outside of lobe material must contribute to the thermal content of the lobes via mixing, which becomes increasingly more important for lower power jets. We do not expect turbulent dissipation to play a significant role here as the turbulent energy content of the lobes is only a small fraction of the total injected energy (see Fig.~\ref{fig:lobe_energy_evo}).

Therefore, considering the bottom row, which shows the total thermal energy injection for all shocks in the vicinity of the jet lobes, we see that for all three jet powers, there is a significant increase in the amount of energy available. Qualitatively, the energy injection rates show similar behaviours to the {\it internal} shocks, although adiabatic compression becomes even more important. The importance of weak shocks (grey band) increases significantly, and thermal energy production can persist beyond $20$~Myr as the lobes continue to expand and interact with the ICM. We also note that at $\sim 20$~Myr, shocks driven by the in-falling substructure discussed previously, intersect the shocks that have travelled furthest from the jet lobes. We attempt to exclude shocks driven by the substructure from our analysis, although this become difficult at later times (light grey shaded region in the figure) when the bottom lobe and substructure drive shocks into each other, which for example likely explains the up-kick in shock energy seen at late times in the fiducial run. 

This being said, even without additional shocks due to this substructure, it is clear that the total jet-driven shocks can explain the required conversion of jet kinetic to thermal energy and after the initial $\sim 5$~Myr also provide additional thermal energy to the ICM. Note that the weak shock contribution becomes increasingly more important at low powers and that the total amount of thermal energy generated via shocks increases with jet power. This analysis paints a picture in which a combination of internal shocks and material passing through the jet-driven bow-shock contribute to the thermal energy budget of the jet lobes, with internal shocks becoming increasingly more important as the jet power increases. Additionally, shocks driven into the ICM by the propagation of the jet and inflation of the jet lobes can contribute to thermal energy in the ICM, with weak shocks being more prominent for low jet powers. 

\begin{figure*}
\psfig{file=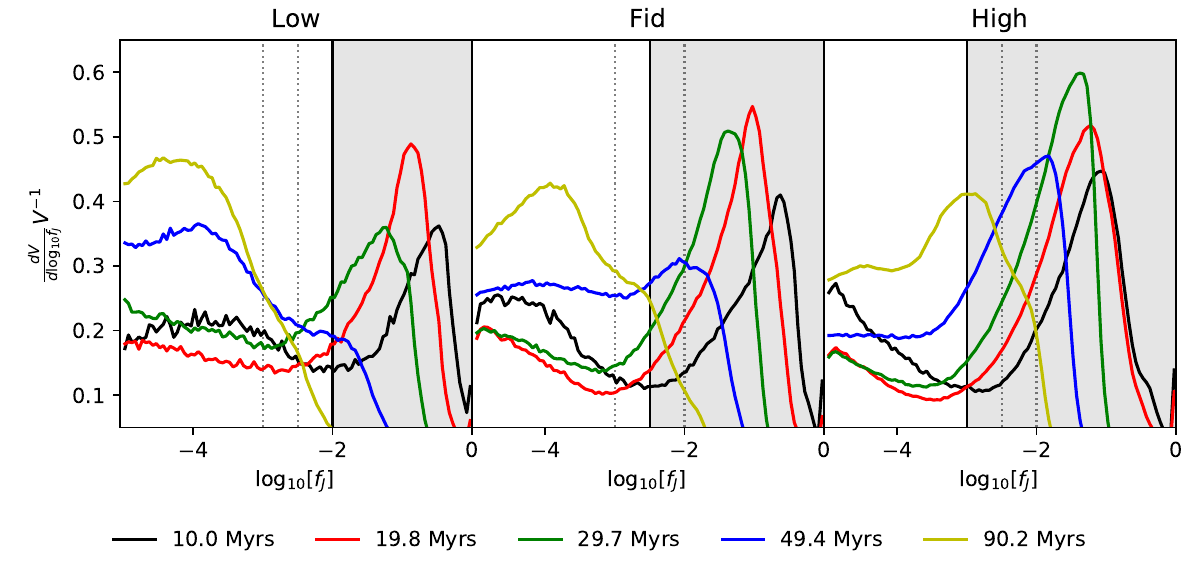,width=1.\textwidth,angle=0}
\caption{Evolution of the jet tracer volume probability distributions with time as a proxy of mixing efficiency. From left to right the panels show the volume probability distribution as a function of $f_{\rm J}$ (and normalised by total volume) for the low, fiducial and high power jet runs, respectively. Each panel contains distributions at five different times and the values of $f_{\rm J}^{\rm thresh}$ are indicated by the vertical lines (solid for the corresponding jet power, dotted for other jet powers). The figure shows that mixing is more effective for lower power jets, where the peak in the distribution moves further to the left (lower jet tracer fraction) and sooner, when compared to higher power jets.}
\label{fig:jet_mixing_distribution}
\end{figure*}

The bottom row of Fig.~\ref{fig:shock_energy_evo} shows that weak shocks are important for injecting thermal energy into the ICM for all jet powers. Without weak shocks, the total shock energy budget at $19.8$~Myr is between $18-27\%$, $33-50\%$, $72-102\%$ of $E_{\rm inj}^{\rm tot}$ for the low, fiducial and high power jets, respectively. When weak shocks are included, we estimate that the equivalent of between $63-94\%$, $92-131\%$ and $122-168\%$ of $E_{\rm inj}^{\rm tot}$ contributes to shock thermal energy production in total for the low, fiducial and high power jet runs, respectively. While we could expect these to exceed $100\%$, given that the same energy can be recycled via multiple shocks resulting in double counting of the same energy\footnote{For example, jet kinetic energy is converted to thermal energy via shocks, which inflates the jet lobes. This thermal energy converts to kinetic energy as the lobes expand, which is then converted back to thermal energy via shocks driven by the lobe expansion.}, the upper values, particularly for the high power jets, seems excessive. Even if we assume an extreme case in which all of the jet kinetic energy is thermalised through shocks and the resulting $PdV$ work done by the inflation of the lobes on the ICM also goes into thermal energy, this sets an upper limit of $\sim 145\%$. In reality, the ``detached'' sound waves seen in Fig.~\ref{fig:mach_projection} would themselves contain a fraction of this energy, and so the true upper limit for pure thermal energy injection via shocks would be somewhat smaller than this. As such, measuring shock thermal energy budgets similar to or exceeding this value would indicate that we are detecting spurious weak shocks/sound waves, capturing shocks from processes other than the jet inflation (e.g. the infalling substructure), and/or pre-existing ICM kinetic energy is additionally being thermalised via shocks at the lobe/ICM interface. The truth is likely to be a combination of these, and as such the values we quote here should be taken as strict upper limits on the shock energy budget.

Taking into account the thermal energy required from shocks to inflate and thermalise the lobes, $E_{\rm lobe}^{\rm req}$, and subtracting it from the total shock thermal energy budget, we find that when weak shocks are included, by $19.8$~Myr, the equivalent of between $0-21\%$, $15-54\%$ and $43-90\%$ of $E_{\rm inj}^{\rm tot}$ for the low, fiducial and high power jet runs, respectively, is potentially available to be communicated to the ICM\footnote{We note that by applying the $f_{\rm J}>10^{-15}$ jet tracer cut when estimating the feedback shock energy budget, this estimate does not include any energy stored in the ``detached'' sound waves seen in Fig.~\ref{fig:mach_projection}.}. We stress, however, that for the reasons discussed above, these values are upper limits, which especially for the high power jets are likely to overestimate the amount of thermal energy injected into the ICM via jet-driven shock. However, a clear picture still emerges showing that higher power jets can contribute more thermal energy to the ICM via shocks as a fraction of their total energy, and as already discussed above, this is facilitated via stronger (high $\mathcal{M}$) shocks. Combining these results provides validation for the picture presented in Section~\ref{sec:icm_heating} in which systems with more powerful jets can retain more thermal energy (as a fraction of injected energy), and conversely, the energy communicated to the ICM via lobe expansion of weaker jets more likely leads to relatively gentle displacement of the ambient medium rather than violently shock heating it.

\subsubsection{Mixing}
\label{sec:mixing}

At late times, mixing has the potential to become a dominant energy transfer mechanism, where it can be comparable to or even exceed our $PdV$ estimates. However, there are some caveats when studying mixing in numerical simulations and trying to untangle physical and numerical effects. While Eulerian codes inherently generate entropy through mixing, they tend to suffer from numerical over-mixing \citep{SpringelArepo2010, MitchellEtAl09, Hopkins15}. On the other hand, an inability to accurately capture fluid instabilities means that mixing is inefficient/suppressed in SPH codes \citep{AgertzEtAl07, DolagEtAl2005, SpringelArepo2010, SijackiEtAl12} unless additional numerical prescriptions are included \citep[e.g.,][]{WadsleyEtAl08, ReadHayField12, ReadEtAl10, Price08}. However, the moving mesh nature of {\sc arepo} means that it can capture mixing without suffering from significant over-mixing \citep{SpringelArepo2010, SijackiEtAl12}. Additionally, through the use of appropriate refinement criteria over-mixing of jet lobes in idealised cluster environments through shredding \citep[e.g.,][]{ReynoldsEtAl05} is also avoided when appropriate resolution is afforded to jet material \citep{BourneSijacki17, WeinbergerEtAl17, EhlertEtal18}.

This being said, an implicit assumption within our simulations is that the fluid within a resolution element is mixed instantaneously, which although results in entropy generation when fluids of different specific entropies combine, neglects physical processes driving mixing below the resolution scale, and could potentially result in shorter timescales than expected physically (with obvious extra levels of complexity arising  when considering mixing of relativistic and thermal plasmas in the presence of magnetic fields). For example, while turbulence and fluid instabilities (e.g. Kelvin-Helmholtz and Reyleigh-Taylor) are captured within our simulations, the physical scale on which the instabilities and turbulence that drive mixing occur are limited by our resolution. Interestingly, however, our previous work found that although improved resolution resulted in reduced mixing between lobes and the ICM, the trend was converging and overall differences in the lobe masses and energy content were small, thus suggesting that at high enough resolution further improvements lead to not very significant changes in the mixing rates \citep{BourneSijacki17}.

Bearing this in mind, in an attempt to illustrate the {\it relative} importance of mixing for different jet powers, in Fig.~\ref{fig:jet_mixing_distribution} we plot the volume probability distribution as a function of jet tracer normalised by the total volume, $dV/(Vd\log_{10}f_{\rm J})$, i.e. the fraction of the total volume in the distribution with a jet fraction between $\log_{10}f_{\rm J}$ and $\log_{10}f_{\rm J}+d\log_{10}f_{\rm J}$, from low to high jet power, left to right, respectively. The distributions are shown at different times by different coloured curves (see legend), the regions of jet tracer space occupied by the lobe material is shown by the grey shaded regions bounded by the solid, vertical black line, while vertical dotted lines indicate the tracer threshold values used for the other two jet powers. 

As jet material mixes with the ambient medium, it is diluted, reducing its value of $f_{\rm J}$. With time the amount of jet material mixing to lower values of $f_{\rm J}$ increases, and a larger fraction of the gas volume is occupied by more dilute jet material driving the peak of the distribution to the left. This behaviour is seen for all three jet powers, with the distributions for the three earliest times remaining within the jet lobe boundary, albeit with the peaks still moving to the left. However, after $\sim 30$~Myr, differences begin to appear; the peak has shifted well to the left for the low power jet by $49.4$~Myr, while the same can be said of the fiducial jet power only at $90.2$~Myr. On the other hand, at $90.2$~Myr, the peak in the high jet power distribution is still on the boundary between jet and non-jet material. Putting aside the differences in jet tracer thresholds used for each jet power run, the late time peaks occur at much smaller values of $f_{\rm J}$ for lower power jets, indicating that the low power jets are more readily mixed. This is in agreement with the findings in Section~\ref{sec:lobe_energetics}. While we expect that the lower power jet lobes are more susceptible to the cluster weather and therefore are more readily mixed, we should also highlight that similar conclusions have been found previously in more idealised environments \citep[e.g.][]{EhlertEtal18}.

\subsection{Impact on cluster properties}
\label{sec:cluster_properties}
\subsubsection{Overview and radial profiles}
\label{sec:cluster_properties_overview}

\begin{figure*}
\psfig{file=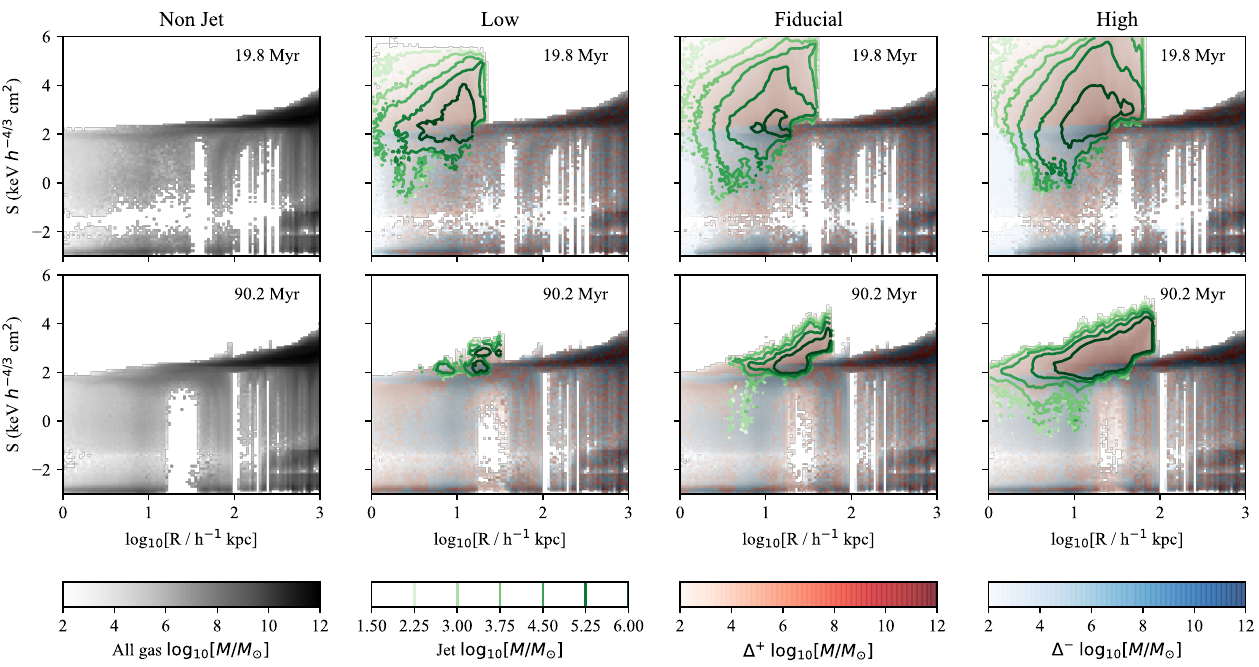,width=1\textwidth,angle=0}
\caption{Two dimensional radial entropy histograms, shown for jets of increasing power from left to right and at $t=19.8$~Myr and $90.2$~Myr in the top and bottom rows, respectively. The total mass-weighted histograms are shown in the greyscale, while residuals from subtracting the non-jet profile from each jet profile are shown in red ($+ve$) and blue ($-ve$). Additionally, the jet lobe content is shown by contours of jet material mass in green. The figure shows that the jets inflate lobes of high entropy gas which move out through the cluster atmosphere and modify the entropy of ambient medium by generally increasing it.}
\label{fig:entropy_profile}
\end{figure*}

To give an overview of how the jets impact the cluster gas at early and late times, Fig.~\ref{fig:entropy_profile} shows 2D radial histograms of the gas entropy (grey) at $19.8$ and $90.2$~Myr in the top and bottom rows, and for the non-jet, low, fiducial and high power jets from left to right, respectively. The red and blue colours in the jet run plots show positive and negative residuals when the non-jet profile is subtracted from the corresponding jet-run profiles. Jet lobe material is shown by the green contour lines. The high entropy jet lobes are visible in the top row, with higher entropy gas existing in the lobes of higher power jets, which heat the gas to higher temperatures and inflate larger, lower density lobes. Considering the residuals in the top row, it is also clear that jets remove material irrespective of entropy at smaller radii but also remove low entropy material at intermediate radii. At later times, high energy jet lobes reach larger radii, but contain typically lower entropy gas compared to younger lobes, while the low power jet lobes appear as only a small bump above the background entropy level. The impact of the jets on the ambient medium can still be seen at late times as coherent regions of red and blue, where the radial entropy distribution of material is different to that of the run with no jet, typically showing that the jets remove mass from low radii and low-to-intermediate entropy regions of phase space. All jet powers cause noticeable increases in the ICM entropy compared to the run with no jet feedback, with the highest power jet resulting in longer-lived features that extend to larger radii.

In using a self-consistent cosmological simulation including sub-grid models for gas radiative cooling and heating, star formation, evolution, enrichment and stellar feedback, our resultant galaxy cluster contains a plethora of features not found in idealised set-ups. Our simulations capture the cluster weather driven by processes such as mergers and cosmic accretion. Additionally, the simulations capture a range of  gas phases that are relevant to galaxy and cluster evolution. In what follows it is useful to consider the impact that the jet feedback has on each of these different phases and how gas transitions between them. We define four key phases (see a phase diagram in Appendix~\ref{app:phase_diagram}) whereby the gas is split into jet material, which we have already defined as non-ISM gas with $f_{\rm J}>f_{\rm J}^{\rm thresh}$, ISM,  which is any gas with densities above $\rho_{\rm crit}=8.5\times 10^{5}$~h$^{2}$~M$_{\odot}$~kpc$^{-3}$ and is considered star-forming, the ICM, which is defined using the temperature-density cut, $T_{\rm keV} > 3\times~10^6\rho^{0.25}_{\rm cgs}$, of \citet{RasiaEtAl12}, and finally the warm phase, which sits between the ICM and ISM phases and is defined as $T_{\rm keV} < 3\times~10^6\rho^{0.25}_{\rm cgs}$ and $\rho < \rho_{\rm crit}$. 

\begin{figure*}
\psfig{file=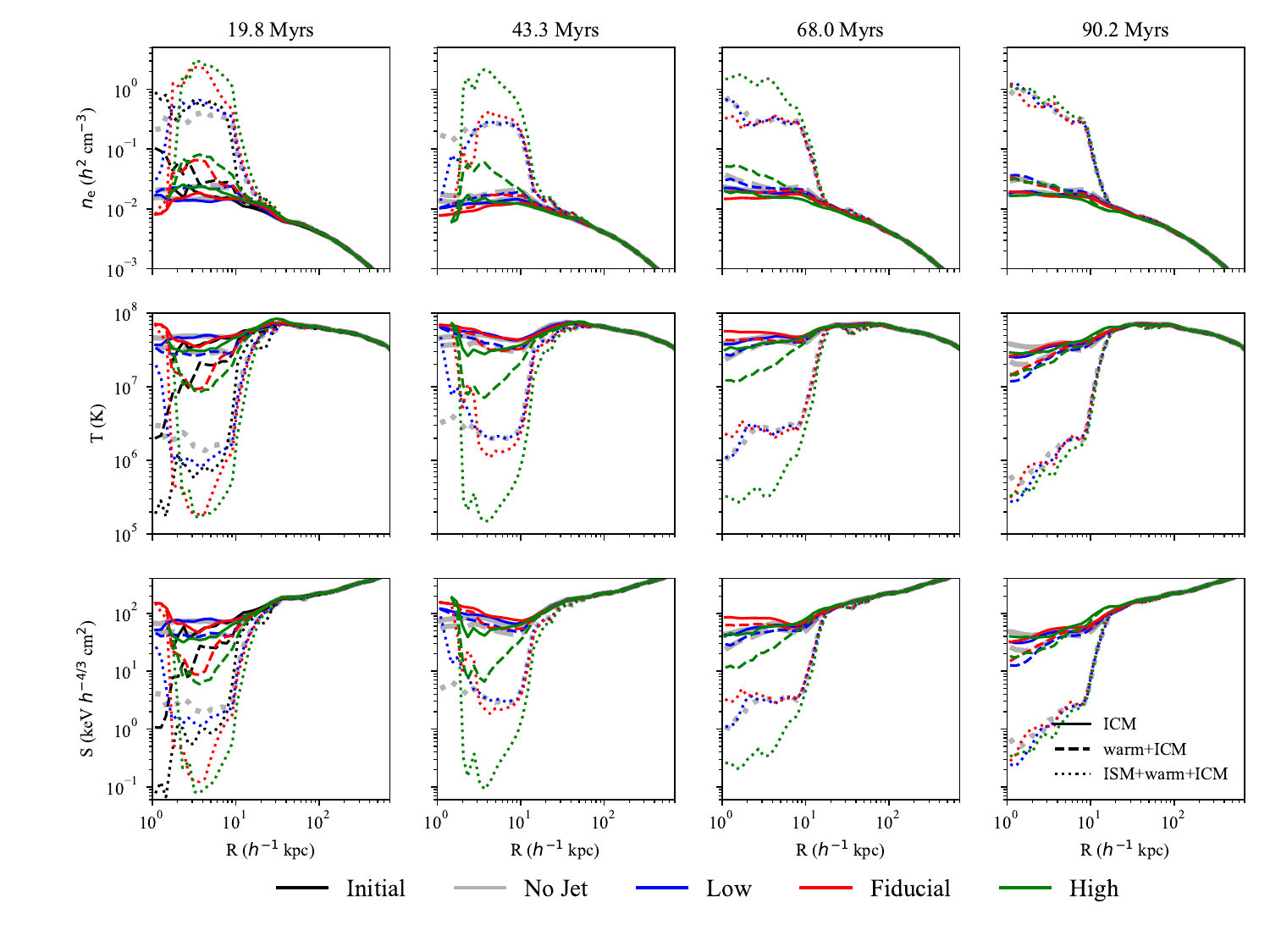,width=1\textwidth,angle=0}
\caption{Radial profiles for density (top), temperature (middle) and entropy (bottom) at different stages of the jet evolution, moving forward in time from left to right. Different line colours correspond to different jet powers, while the different line styles correspond to different gas phase combinations, namely ICM (solid), ICM+Warm (dashed) and ICM+Warm+ISM (dotted). Note that jet material is neglected in all profiles. Although differences can be seen in the profiles due to the jet action, these are largely confined to earlier times and for higher power jets. The ICM profile varies very little between different jet powers, while all profiles look very similar by $90.2$~Myr.}
\label{fig:radial_profiles}
\end{figure*}

To provide a general overview of the impact of the jet on different gas phases cluster radial profiles are shown in Fig.~\ref{fig:radial_profiles}. The figure is arranged as follows: the top, middle and bottom rows show the radial density (mass-weighted), temperature (volume-weighted) and entropy profiles, respectively, while from left-to-right the columns show profiles calculated at different times during the jet evolution. The different jet powers are shown in different colours (as shown in the legend at the bottom of the figure) while different line styles indicate different gas-phase combinations. Specifically, the solid lines show pure ICM gas (ICM profile), the dashed lines show the combined ICM and warm phase profiles ({\bf W}arm-{\bf H}ot profile, WH), while the dotted lines show the combined profile of the ISM, warm and ICM phase ({\bf C}old-{\bf W}arm-{\bf H}ot profile, CWH), noting that the jet material is excluded from all profiles. Additionally, in the first column, we show the initial cluster profile (before jet switches on) with the black lines. Before discussing these profiles we want to stress two key points: firstly, the initial cluster profiles are the result of sub-grid models typical of cosmological simulations and not a reflection of the new jet feedback model we present here, and secondly, the jet feedback we implement assumes a fixed jet power and lifetime and is in no way coupled to the cluster properties, as such these profiles are not presented as strict comparisons to observations but rather as an analysis tool for understanding the cluster heating process.

Firstly, a brief look at all of the profiles shows that beyond the central $\sim 30$~h$^{-1}$~kpc, the differences in profiles become negligible. This is to be expected as the total injected energy is equal to the initial ICM thermal energy content within radii of $\sim 10$~h$^{-1}$~kpc, $18$~h$^{-1}$~kpc, and $30$~h$^{-1}$~kpc, for the low, fiducial and high power jet runs, respectively. This means that any energy injected into the ICM at larger distances very quickly becomes small compared to the total ICM energy contained within this region. It is also worth pointing out that at these distances the profiles are dominated by the ICM. Considering the profiles just before the jets turn off, at $19.8$~Myr, jets of all powers have transferred $\sim 50\%$ of the jet energy to the local environment and all present differences in the profiles compared to the run with no jet. Specifically, the jets remove gas, reduce the density and heat all phases within the central $1-2$~kpc region, which becomes dominated by jet material (not-shown on this plot). 

Beyond this central region, the CWH and WH profiles show large density enhancements as well as decreases in temperature and entropy when compared to the non-jet run, especially in the case of the fiducial and high power jets, while the ICM profiles show much weaker reactions to the jet action with mild density enhancements and temperature/entropy decreases for the fiducial and high power jets only, with the ICM in the low power jet run showing almost no difference when compared to the non-jet run. The differences between the jet/non-jet run CWH and WH profiles can be driven by two processes: {\it i)} the compression of gas as the inflation of the jet lobes pushes it out to larger radii and/or {\it ii)} the removal of hot ICM and/or warm phase gas. The former process by construction increases the density and temperature of the gas (although ultimately the increased density results in shorter radiative cooling times, with gas temperatures subsequently falling), while the latter process means that the cooler phases make a more significant contribution to average profile quantities giving the appearance of gas densities/temperatures increasing/decreasing. As we show below in Section~\ref{sec:diff_phases}, while the jet does compress all gas phases during the inflation (although this peaks before $20$~Myr), a significant amount of hot ICM gas is also cleared out in the fiducial and high jet power runs and so it is a combination of these two processes which drive the differences in the radial profiles.

Considering only the CWH and WH profiles, at later times, after the jet action has halted, while the high power jet run still exhibits differences, the lower power jet profiles become increasingly more similar to those of the non-jet run. Specifically, at $43.3$~Myr, all jet-powers still show significant differences within the central $\sim 2$~kpc in the CWH profiles, although beyond this both the low and fiducial jet power runs are very similar to the non-jet run and only the high power jet shows large differences for both the CWH and WH profiles. By $68$~Myr, the differences are even smaller, the low and fiducial power jet run profiles show only minor differences to the non-jet run profiles for all radii, while the high power jet profiles only show a significant difference in the CWH profiles. Finally, by $90.2$~Myr even the high power jet run shows only very small differences compared to the non-jet run.

Importantly, if we consider the ICM entropy profiles, we see that the low power jet results in a mean entropy excess in the cluster core even at $43.3$~Myr, while an excess still exists in the $68$~Myr profile for the fiducial jet power and the high power jet run exhibits a not insignificant bump in the entropy profile out to $\sim 20$~kpc at $90.2$~Myr when compared to the run without a jet. The jet feedback increases the mean ICM entropy, which in the case of the high power jet persists long after the jet has switched off. Comparing to Fig.~\ref{fig:icm_energy_evo}, we find that the mean entropy excess appears at smaller radii than the peaks in $\delta U/U$. However, this can be understood by noting that although on average higher entropy gas is seen at smaller radii in Fig.~\ref{fig:radial_profiles}, there is also a mass deficit at small radii (but a mass excess at larger radii), that at least in part, explains why the $\delta U/U$ peak does not match the location of the higher mean entropy.
 
However, if we were to take each radial profile in isolation the system evolves such that it {\it relaxes} back to a state that exhibits very few ``scars'' of the jet feedback. As we discuss further in Section~\ref{sec:diff_phases}, in the cluster core, the ISM gas in the jet runs relaxes somewhat such that the mean density and temperature become closer to those measured in the non-jet run, while ICM material that was initially expelled by the jet action begins to repopulate the cluster core, which combined contributes to the radial profiles becoming more similar at later times. Additionally, within the ICM and particularly at larger radii where it dominates the mass and energy budget, it would require a significant energy injection to make a long-lived and noticeable difference to the ambient density and temperature. This implies that in reality, the jet feedback is quite subtle, meaning that even though the cluster retains a significant fraction of the injected jet energy (see Fig.~\ref{fig:icm_energy_evo}), its impact on the ICM radial profiles become less apparent with time as the energy is distributed over a larger volume and converted into potential energy.

\subsubsection{How different phases in the cluster core evolve}

\label{sec:diff_phases}
\begin{figure*}
\psfig{file=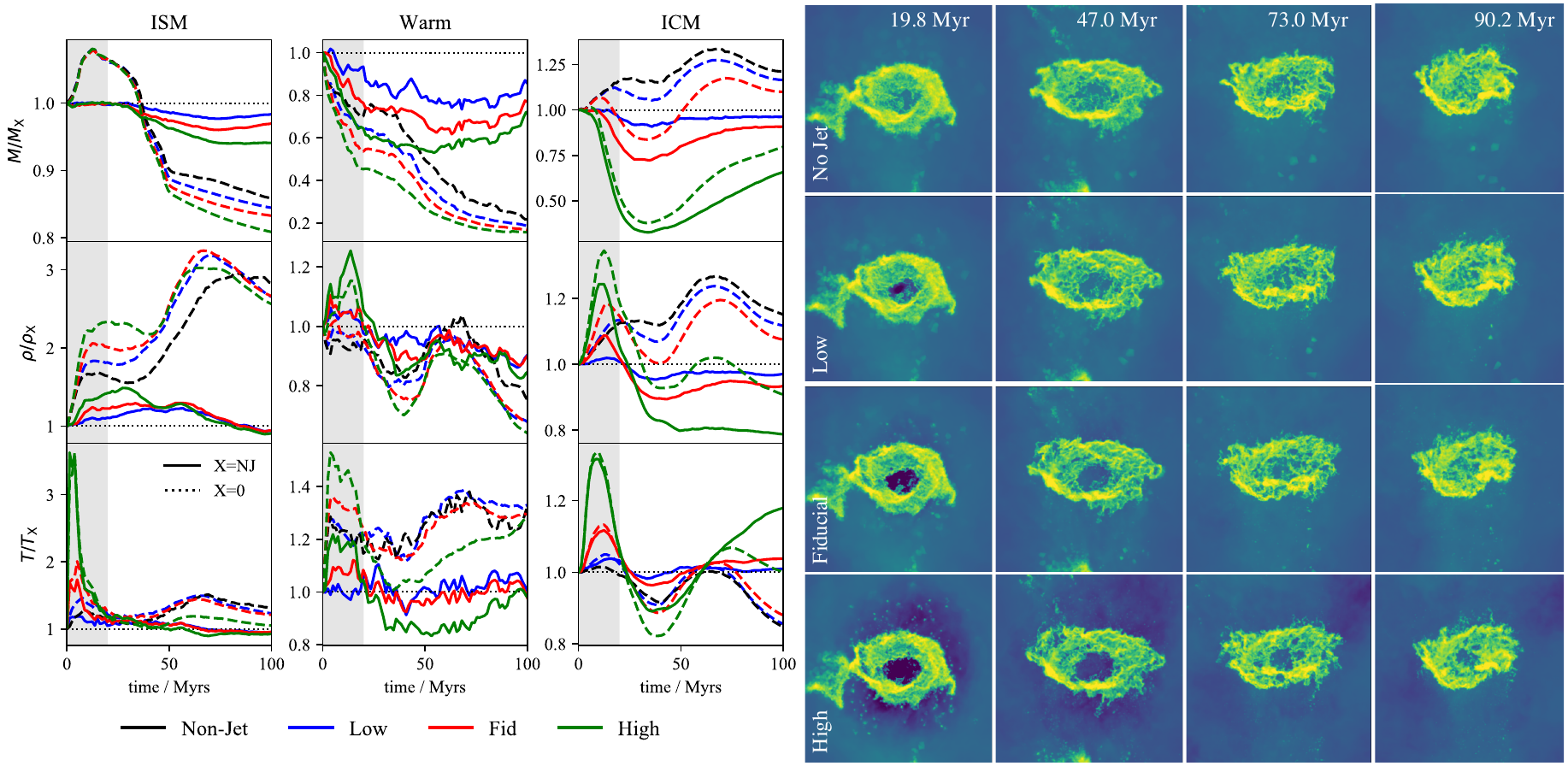,width=1.\textwidth,angle=0}
\caption{Evolution of different gas phases within the centre of the cluster. On the {\it left hand side} we show quantitatively how different gas phases; ISM (left), warm (middle) and ICM (right), evolve with time within the central $20$~h$^{-1}$~kpc of the cluster. We consider how the total mass (top), mean density (middle) and mean temperature (bottom) evolve with respect to the IC (dashed) and the non-jet run (solid curves). On the {\it right hand side} we show column density projections (through boxes of $40$~h$^{-1}$~kpc on a side) for each of the four runs, increasing with power from top to bottom and evolving in time from left to right, respectively. The figure illustrates how different power jets impact the phases early on, but at later times, the system is able to some extent relax back to similar states, most noticeably for the cold phase and least for the ICM.}
\label{fig:phase_mass_evo}
\end{figure*}

We explore the impact of the jet feedback in the cluster centre ($r\leq 20$~h$^{-1}$~kpc) in more detail in Fig.~\ref{fig:phase_mass_evo}. On the left-hand side we plot, the evolution of the ISM (left column), warm phase (middle column) and ICM (right column) total mass (top row), mean density (middle row) and mean temperature (bottom row) normalised by either the initial value at $t=0$ (dashed) or the value in the non-jet run at the same time (solid curves). The non-jet, low, fiducial and high jet power runs are shown by the black, blue, red and green lines, respectively. Additionally, on the right-hand side, we show column density projections of the central region for all four runs increasing in power from no jets to high power jets from top to bottom, respectively, and evolving in time from left to right. Due to the nature of our simulations, which includes processes other than those solely driven by AGN feedback, even the non-jet run (black lines) shows an evolution in all of the properties considered, however, we see that the jet feedback introduces differences in how these properties evolve. 

First considering the ISM, which dominates the central gas mass budget ($79\%$), we note that the structure evolves significantly even in the non-jet run, highlighting that processes including star formation and associated feedback as well as dynamical interactions due to passing substructures can have an impact on the cold gas properties. At early times the central ISM mass increases primarily due to a substructure, which can be seen to the left of the BCG at $19.8$~Myr in the column density projections, entering the region. While the ISM mass remains similar for all simulations, it is clear that the distribution of this material is different. In particular, the jet simultaneously excavates a hole in the centre of and expels material from the cold disc\footnote{Note that the low accretion rates assumed for our jets mean that the central cavity in the cold disc forms through the inflation of the jet opposed to the accretion of cold material.}, with the effect being more extreme for more powerful jets. There are also differences in the mean density and temperature of the ISM. Lobe inflation compresses gas in the central disc, increasing its density and hence shifting warm gas into the ISM phase as well as ``pushing'' ISM gas along the effective EoS. This results in early peaks in the mean ISM temperature, especially for the fiducial and high power jets, which subsequently drop as the gas cools. This means that during inflation, while the jet does not destroy significant amounts of cold gas, it can redistribute it and change its properties. As we discuss later in Section~\ref{sec:star_formation}, the increase in density impacts the SFRs found in the BCG. 

After $\sim 14$~Myr, the total mass begins to slowly decline in all runs as a substructure begins to leave the region, and the decline steepens after $\sim 30-35$~Myr as the object finally leaves the central region. At around the same time, small differences in the ISM mass between the runs start to appear and while by $\sim 40$~Myr, the central densities in the three jet runs become similar in value and evolution, they appear to be systematically offset to higher values and peak earlier than the non-jet run. By this time, other substructures have left the central region and differences in the total ISM mass for all four runs become more obvious, with cold material expelled by the jet continuing to leave the central region. However, by late times we see that while there are small differences in the total ISM mass, the mean ICM density and temperatures are similar for all jet and non-jet simulations and looking at the column density projections we see that despite the perturbations caused by the jet the cold gas appear to have a very similar structure, which, hides any jet induced ``scars''.

The warm phase, which contributes the least to the total gas mass ($\sim 5\%$), exhibits a continuous drop in mass throughout all four simulations, as it is either heated and joins the ICM or cools onto the ISM. While these processes occur in the non-jet run, the action of the jet seems to enhance them, with jet simulations retaining less warm phase material, as it is either heated or compressed by the jet and transitions to one of the other phases. The density and temperature of the warm phase fluctuate over time, but as in the case of the ISM, there is a clear compression during the jet feedback phase seen as a density and temperature increase, with the increase being larger for higher power jets. 

Finally, considering the ICM, in general, the non-jet run exhibits an increase in mass in the central region up until $\sim 70$~Myr, at which point it begins to decline. In terms of the total mass content, which comprises $16\%$ of the central gas mass initially, we can see that different jet powers can have very different effects. While the low power jet results in a small reduction in the mass of ICM compared to the non-jet run, the fiducial jet results in a decrease in the total ICM material for early times, while the high power jet produces a long term reduction in the amount of ICM in the cluster centre. This is a direct result of the jets and the lobes they inflate; displacing and clearing out the hot ICM material that once resided at the location of the lobes and can be seen in the early column density projections that show a reduction in ICM material in the central region surrounding the disc. The action of ``sweeping-up'' the ICM compresses it, resulting in clearly defined density and temperature peaks seen during the first $\sim 20$~Myr, as well as the production of the X-ray bright rims shown in Fig.~\ref{fig:overview}. We additionally note that this compression results in shorter gas cooling times within the higher density gas.

However, once the compressed material leaves the central region there is a drop in the mean density and temperature, which continues for a short time after the jet switches off, with differences compared to the non-jet run, especially for the high power jet, continuing to exist. We also note that shortly after the jet action stops the ICM starts to replenish with the differences in mass content between the runs becoming smaller with time. Additionally, the increased ICM cooling rates observed during the lobe inflation decrease after the jet switches off. We find that the changes to ICM properties induced by the jet action lead to ICM cooling rates becoming lower than those in the non-jet run, even out to $\sim 100$~h$^{-1}$~kpc in the case of the high power jet, which will hinder future gas cooling inflows. Overall, this figure paints a picture in which the general evolution of the central ISM is largely determined by processes other than the jet feedback, at least on the timescale of our simulations. However, the jet can moderate - either increase or reduce - variations in the properties of all phases to varying degrees, and particularly in the case of the higher power jets reduce the ICM and warm phase gas content. This has the potential to affect the reservoir of hot gas that could cool in the future and can also have additional implications that result from the non-linear evolution of the system, for example through the impact of AGN feedback on star formation, as we discuss next. 

\subsubsection{Star formation}
\label{sec:star_formation}
\begin{figure}
\psfig{file=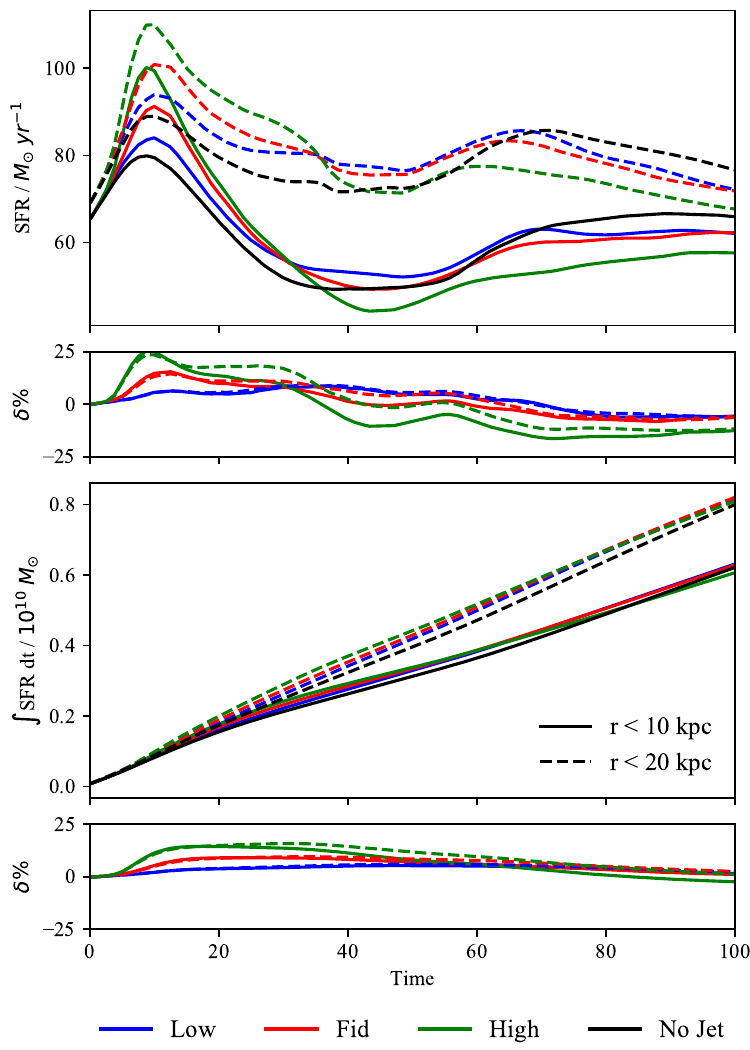,width=0.5\textwidth,angle=0}
\caption{Star formation within the centre of the cluster, shown for $r<10$~h$^{-1}$~kpc (solid) and $r<20$~h$^{-1}$~kpc (dashed). From top to bottom, respectively, the panels show the total SFR within the corresponding regions, the fractional difference in SFR between jet and non-jet runs, the integrated SFR and the fractional difference in the integrated SFR between jet and non-jet runs. During the inflation phase jets can increase the central SFR through the compression of gas, but at later times the SFRs subside and the net difference in the integrated stellar mass is similar between all runs.}
\label{fig:sfr_evo}
\end{figure}

As shown in Section~\ref{sec:diff_phases}, the jet can compress gas to higher densities. The resulting increased density in the ISM will have an impact on the SFR of the central BCG, which we now explore. Fig.~\ref{fig:sfr_evo} is split into four panels that from top to bottom show the SFR, the fractional difference in SFRs between jet and non-jet runs, the integrated stellar mass and the fractional difference in integrated stellar mass between jet and non-jet runs, respectively. The different simulation runs are shown by different colours, as indicated by the legend, while the solid and dashed lines show values within $10$~h$^{-1}$~kpc and $20$~h$^{-1}$~kpc, respectively. 

In line with the increase in density of the cold ISM seen in all runs during the first $\sim 10$~Myr, the SFR also increases. While the low power jet only provides a small increase with respect to the non-jet run, the difference becomes larger with increasing jet power, with the high power jets resulting in a $\sim 25\%$ increase over the non-jet run. This is due to the higher power jets being able to compress gas to higher densities, which results in higher SFRs, with similar findings found by some previous works \citep[e.g.,][]{MukherjeeEtAl18, ValentiniEtAl20,ZubovasBourne17}. Despite the increased SFRs during the jet outburst, the total integrated stellar masses after $100$~Myr are similar between all runs. This is because even though the SFRs remain elevated in the jet runs even after the jets deactivate, eventually they drop below the SFR in the non-jet run, firstly for the high power jets, followed later by the fiducial and then the low power jets. Differences in SFR will depend on both the amount of cold gas available and the properties of that gas. Considering the left-hand column in Fig.~\ref{fig:phase_mass_evo}, we see that while there are corresponding peaks in SFR and mean ISM density, the size of the peaks is not correlated, i.e., the strong early SFR peak does not correspond to the highest density peak. As such, we expect that the late time differences in SFR between different runs are governed primarily by the amount of cold ISM available and as such that the behaviour which drives the differences in the mass of ISM in the central region (top left panel, Fig.~\ref{fig:phase_mass_evo}), is also responsible for the decrease in SFR. 

\section{Discussion}
\label{sec:discussion}

The art of simulating galaxy cluster populations has come into its own in
recent years, with a number of groups performing suites of cosmological-zoom
simulations that include a range of different cluster masses
\citep{LeBrunEtAl14,BarnesEtAl17,BarnesEtAl17b,BarnesEtAl18,McCarthyEtAl17,BaheEtAl17,
  HahnEtAl17,RasiaEtAl15, HendenEtAl18, HendenEtAl19, HendenEtAl20}. These
works have made great progress in being able to reproduce many observed galaxy cluster properties, however, matching thermodynamic profiles, particularly in
the cluster cores, and explaining the cool-core/non-cool-core (CC/NCC)
dichotomy remains a challenge. For example, overly effective AGN feedback in
the Illustris simulations resulted in too much gas being ejected from groups
and low mass clusters \citep{GenelEtAl14}. This issue was resolved for
Illustris-TNG, although the CC fraction and its evolution do not match
observations due to differences between simulated and observed gas fraction
profiles \citep{BarnesEtAl18}, suggesting that the AGN feedback model needs
modification. Another symptom of overly efficient AGN feedback was evident in
both C-Eagle \citep{BarnesEtAl17} and FABLE \citep{HendenEtAl18}, which while
being able to match multiple observables, found overheating in the cluster
centres. Finally, while Rhapsody-g simulations could reproduce the CC/NCC
dichotomy, it was at the expense of matching observed X-ray luminosities
\citep{HahnEtAl17}.

The work we have presented here attempts to improve our theoretical
understanding of AGN jet feedback by bridging the gap between jet injection
and galaxy cluster scales. The focus hence shifts to model  observational
signatures and behaviour of the jet itself as opposed to its impact on the
cluster environment, which we hope to arise naturally from a more physically
motivated feedback model. It is important to highlight that while several
cosmological works invoke jet mode feedback, this typically mimics the effect
of the jet feedback, for example by adding hot bubbles by hand
\citep[e.g.,][]{SijackiEtAl06a,SijackiEtAl07,SijackiEtAl2015}, or by injecting
bipolar kinetic outflows
\citep[e.g.,][]{DuboisEtAl10,DuboisEtAl12,DaveEtAl19}. The models are 
typically (and by necessity) simplified and not designed to track the lobe
inflation and their subsequent evolution in detail.

There are a limited number of
previous studies that similar to this work attempt to capture the lobe inflation
and evolution in a cosmological environment
\citep[e.g.,][]{Heinz2005,MendygralEtAl12,MorsonyEtAl2010}. Here we expand
upon these studies by including additional physical models (such as
radiative cooling, star formation and associated feedback), by forcing very high
resolution in the jet injection region and {\it crucially} by adopting a suite of
refinement techniques that ensure that the high resolution ($\sim
120$~h$^{-1}$~pc) tracks the jet lobes at all times, to capture the
details of the lobe inflation process and its subsequent interaction with the
ICM. Importantly, we find in common with these previous studies \citep[see
  also][]{SijackiEtAl08, BourneSijacki17} that cluster weather, due to the
pre-existing turbulence, bulk motions, and orbiting substructures can play a
very significant role in redistributing and/or disrupting jet lobes.   

In idealised simulations of jet feedback in hydrostatic atmospheres, jet lobes
inflate and rise along the jet axis. Such simulations result in low density
channels along one axis through which feedback energy can easily escape and
hence couple poorly with the ICM \citep{VernaleoReynolds06}. Previously this issue has been overcome in simulations by including precession
\citep{Falceta-GoncalvesEtAl10, LiBryan14, YangReynolds16Hydro,
  YangReynolds16ThermalConduction} or rapid re-orientation of the jet axis by
hand \citep{CieloEtAl18} to isotropise and increase the coupling
between the jet energy and ICM. However, when cluster weather is included, as
found here, the jet lobes can be naturally displaced from their original
trajectory by bulk motions in the ICM \citep[see also,][]{BourneSijacki17,
  SijackiEtAl08, MorsonyEtAl2010}. Given the expected difficulty in driving
precession of the jets from massive BHs by accretion \citep{NixonKing13},
cluster weather provides a possible explanation for observations of successive
generations of jet lobes being misaligned in some galaxy clusters
\citep{DunnEtAl06, BabulEtAl13}.

Jet lobes in systems with cluster weather are effectively mixed with the ICM
\citep[see also,][]{BourneSijacki17, Heinz2005, MorsonyEtAl2010}. We note that
some previous works have suggested that 
mixing of hot lobe material with the ICM through jet-driven instabilities is
an effective method of heating the ICM \citep{HillelSoker16, HillelSoker17},
even in a hydro-static environment. However, we stress that here we are
emphasising the importance of cluster weather-driven mixing, noting that at
sufficiently high resolution and by employing appropriate refinement criteria,
in \citet{BourneSijacki17} we found instability driven mixing to be largely
unimportant in hydrostatic environments, even in the absence of magnetic
fields \citep[see also,][]{WeinbergerEtAl17,EhlertEtAl19}. We further note
that while our simulations are performed without magnetic fields, which have
been shown to affect lobe dynamics, and inhibit small-scale instabilities
along the lobe-ICM interface that facilitates mixing \citep{DursiPfrommer08,
  WeinbergerEtAl17,EnglishEtAl16}, we expect that the drastic nature of the
cluster weather on lobes (particularly the bottom lobes) would still have a
strong impact and drive mixing even when magnetic fields are included.

An important (and related) point to consider is the potential relationship
between the dynamical state of the ICM and whether or not a cluster is a CC or
NCC. Observations, such as the REXCESS sample, show that there is a lack of
cool-cores in dynamically disturbed systems \citep{PrattEtAl10}, while
\citet{MahdaviEtAl13} found that while NCC clusters exhibit an X-ray mass bias
compared to weak lensing mass measurements, this is not the case for CC
clusters, suggesting that CCs are dynamically relaxed. If, as we have found
here, cluster weather can disrupt jet lobes and aid in the process of
heating the ICM, then we could expect X-ray cavities to be more common in CC
clusters, as is the case from observations
\citep{Burns90,DunnEtAl05,DunnFabian06,DunnFabian08,McNamara2007,Sun09,Fabian12}. The common occurrence of lobes in CC clusters has often been attributed to a link
between cooling, BH feeding and subsequent feedback, however, there may also
be an observational bias in that lobes are more likely to survive and hence are preferentially observed in dynamically quiet environments.

Additionally, the potentially lower efficacy of feedback in CCs may limit its
role in disrupting and converting them to NCCs. Even in the simulations
presented here in which cluster weather helps liberate feedback energy, it
does not have a significant impact on the cold gas content of our
cluster. However, the feedback may instead be able to aid in preventing CCs
forming in the first place. For example \citet{McCarthyEtAl08} found that
early AGN heating can indeed inhibit the formation of CCs. Alternatively,
numerous simulations have found that it is mergers, and not AGN feedback, that
disrupts CCs \citep[e.g.,][]{BurnsEtAl08, PooleEtAl08,RasiaEtAl15, HahnEtAl17,
  ChadayammuriEtAl20}, with, \citet{HahnEtAl17} and \citet{ChadayammuriEtAl20}
finding that, similar to the results presented here, AGN feedback is
ineffective at significantly altering/removing the cold gas content of the
cluster core. This being said it is also worth noting that
\citep{BarnesEtAl18} found from the Illustris TNG cosmological simulations that
there is no difference in the dynamical state of CCs versus NCCs, which would instead suggest mergers play a limited role in converting CCs to NCCs. These
differing conclusions illustrate the need for future simulations to combine
both realistic cluster environments with physically motivated, high resolution
AGN feedback models.

The simulations presented here consider a single cluster at a single
epoch. This has allowed us to garner insights into how AGN feedback can heat
the ICM and impact different gas phases within objects of this kind. However,
for a full understanding of the role of AGN feedback in shaping galaxy
clusters it will be necessary to consider clusters with a wide range of
properties over the whole of cosmic history to understand how the ICM
forms and evolves; the development of the entropy excess
\citep{VoitEtAl05,SunEtAl09}, turbulence/cluster weather
\citep{VazzaEtAl17,PintoEtAl2015,Hitomi2016,BourneSijacki17}, and magnetic
fields, and how each of these impact the AGN feedback process in both CC and
NCC systems (see discussion above). This will also require the inclusion of a
self-consistent accretion and feedback model that has not so far been included
in our simulations to determine both the jet direction and production
efficiency and close the feedback loop between large scale gas inflows and
central AGN heating. 

Additionally, as with many large scale cosmological simulations, it is not possible to include all physical processes due to a combination of
the large dynamic range of the systems considered, and current computational
or physical limitations. Specifically, we have neglected MHD effects, which
have been shown to inhibit instability driven mixing of jet lobes in idealised
simulations and impact lobe dynamics
\citep[e.g.][]{DursiPfrommer08,EnglishEtAl16,EhlertEtal18,WeinbergerEtAl17},
although these effects are seemingly negated in a large part by implementing
sufficient resolution and refinement criteria \citep{BourneSijacki17,
  EhlertEtal18}. However, given the growing number of cosmological simulations
that successfully include the effects of magnetic fields
\citep[e.g.,][]{DolagEtAl2002,MarinacciEtAl15,MarinacciEtAl18,Martin-AlvarezEtAl18,KatzEtAl19}, it is timely to run a comparison suite in future works, in particular, to check if the cluster weather aided mixing is as prominent when magnetic fields are included. Neglecting magnetic fields additionally means that thermal conduction is not included in these runs, which while adding additional levels of complexity to simulations, has been shown to play a potentially important role in solving the cooling flow problem \citep[e.g.][]{YangReynolds16ThermalConduction,
        VoigtFabian04,RuszkowskiOh10,RuszkowskiOh11,BogdanovicEtAl09,KannanEtAl17}. 

A further simplification we adopted is in modelling the hot lobes as a
non-relativistic thermal gas ($\gamma = 5/3$) in which the electrons and protons share a single
temperature. The exact nature of the lobe plasma is still uncertain
\citep[e.g.,][]{DunnEtAl06b,BirzanEtAl08,CrostonEtAl08,CrostonHardcastle14,KangEtAl14,KawakatuEtAl16,MazzottaEtAl02,SchmidtEtAl02,DunnEtAl05},
and in reality, the two components may not be in thermal equilibrium and have
different temperatures, their distributions may be non-thermal. Uncertainties surrounding how CR energy is communicated to the ICM via streaming and diffusion mean that it is unclear what fraction of the CR energy is actually transferred to the ICM and on what timescales. Given that we do not include any models for CRs with our simulations, the mixing rates we present may not map directly to those expected in real clusters, but rather provide a qualitative mechanism by which cluster weather could enhance mixing of lobe material. In fact, some works
have studied the effect of different lobe content, for example by including
the CR component
\citep[e.g.,][]{SijackiEtAl08,WeinbergerEtAl17,EhlertEtAl19,YangEtAl19},
finding that CRs can impact lobe dynamics, with the additional CR pressure
resulting in more buoyancy and easier uplift of ICM gas, and produce
``rounder'' bubbles compared to kinetically injected jets that are more
similar to observed X-ray cavities, suggesting that CRs could have a role
to play in ICM evolution. 

Finally, we do not include a physical viscosity model and hence any viscous
dissipation occurring is purely numerical. As discussed in
Section~\ref{sec:shocks}, sound waves could contribute to heating the ICM
\citep[e.g.,][]{FabianEtAl05,SandersEtAl07,BambicEtAl19,FabianEtAl17} and as
such a full understanding of where this occurs and at what rate would depend
on the physical viscosity of the ICM \citep{RuszkowskiEtAl04, SijackiEtAl06b,
  ZweibelEtAl18}, which itself is still rather uncertain. However, given that
{\sc arepo} has very low numerical viscosity, sound waves are expected to be
long lived. As such, the fact that the weak shocks/sound waves in the low
power jet run so readily disappear suggests that the cluster weather can at least partially disrupt them, as was also found in
\citet{BourneSijacki17}. Therefore, transporting energy over a large volume via sound waves may only be viable in relatively quiet cluster atmospheres or with sufficiently high jet powers. 

\section{Summary}
\label{sec:summary}
We have presented results from simulations of very high resolution jets launched into a fully cosmologically evolved galaxy cluster. The work builds on that of \citet{BourneEtAl19} by extending the range of jet powers studied and performing in-depth analysis to investigate lobe inflation and dynamics, as well as the jet-lobe energetics and how the jet energy is communicated to and impacts the ICM. A simple picture arises in which rather gentle feedback is able to heat the ICM via a combination of weak shocks and sound waves during lobe inflation, while cluster weather subsequently plays a critical role to effectively liberate lobe energy and transfer it to the ICM. The exact details depend on jet power and our main findings are:
\begin{itemize}
    \item {With appropriate refinement criteria that provide sufficient resolution in the injection region as well as crucially in the jet lobes, it is possible to inflate low density, high temperature lobes within a full cosmological zoom simulation of a galaxy cluster. The lobe properties result in mock X-ray and radio appearances that compare favourably to observations of jet lobes in galaxy clusters.}
    \item{Analysis of lobe energetics show that irrespective of the jet power, the lobes retain $\sim$half of the injected energy during the lobe inflation phase, with the majority of the ``lost'' energy going into $PdV$ work done by the expanding lobes. This is consistent with previous high resolution simulations of lobe inflation in idealised clusters.}
    \item{Shock heating plays a crucial role in converting the jet kinetic energy to thermal energy to inflate jet lobes. Heating by both internal shocks and the bow shock is necessary to explain the lobe energetics, although their relative importance changes with jet power, with internal shocks becoming progressively more important for higher power jets.}
    \item{Shock heating also contributes to ICM heating, with higher power jets providing a larger fraction of the injected jet energy to the ICM via shocks. This being said, weak shocks make up a large portion of the heating (especially for lower power jets), and while ICM Mach numbers increase with jet power, even for the highest power jet explored they rarely exceed $\mathcal{M}=3$, consistent with observations of jet driven shocks in galaxy clusters.}
    \item{Cluster weather displaces jet lobes from their original trajectory, aids their disruption and drives mixing of hot lobe material with the ICM, especially after the jet has switched off. Mixing is more effective for lower power jets which are more susceptible to the cluster weather.}
    \item{While a significant fraction of the jet energy ends up as potential energy of gas in the ICM, we find that systems with higher power jets retain a higher fraction as thermal energy while low power jets retain a larger fraction as kinetic energy. This is consistent with the picture that higher power jet lobes, which are more ``explosive'', convert a larger fraction of kinetic to thermal energy via shocks during more rapid lobe inflation, while lower power jet lobes are more likely to simply displace material.}
    \item{Overall the impact on cluster properties is subtle. During lobe inflation the jets can compress cold, dense ISM gas, and enhance instantaneous SFRs, although this results in only small differences for the long-term integrated stellar mass. However, the jet feedback is able to heat the ICM and gently raise the mean core entropy even long after the jet switches off, with energy in excess of that in the non-jet run being comparable to the total injected jet energy. Given that $\simgt 50\%$ of the injected energy ultimately ends up as gravitational potential energy of the system, differences in the radial ICM profiles appear small at later times as the energy is redistributed and the cluster core relaxes. This is consistent with observations of cluster radial profiles that do not show sharp variations even during AGN outbursts, pointing to feedback needing to be a gentle process.}
\end{itemize}

\section*{Acknowledgments} 
MAB and DS acknowledge support by
the ERC starting grant 638707 ``BHs and their host galaxies:
co-evolution across cosmic time.'' DS further acknowledges support
from the STFC. This research used: The Cambridge Service for Data Driven Discovery (CSD3), part of which is operated by the University of Cambridge Research Computing on behalf of the STFC DiRAC HPC Facility (www.dirac.ac.uk). The DiRAC component of CSD3 was funded by BEIS capital funding via STFC capital grants ST/P002307/1 and ST/R002452/1 and STFC operations grant ST/R00689X/1. The DiRAC@Durham facility managed by the Institute for Computational Cosmology on behalf of DiRAC. The equipment was funded by BEIS capital funding via STFC capital grants ST/P002293/1 and ST/R002371/1, Durham University and STFC operations grant ST/R000832/1. DiRAC is part of the National e-Infrastructure.

\section*{Data availability}
The data underlying this article will be shared on reasonable request to the corresponding author.

\bibliographystyle{mnras}

\bibliography{ArepoJet}

\appendix
\section{Phase Diagram}
\label{app:phase_diagram}
\begin{figure}
\psfig{file=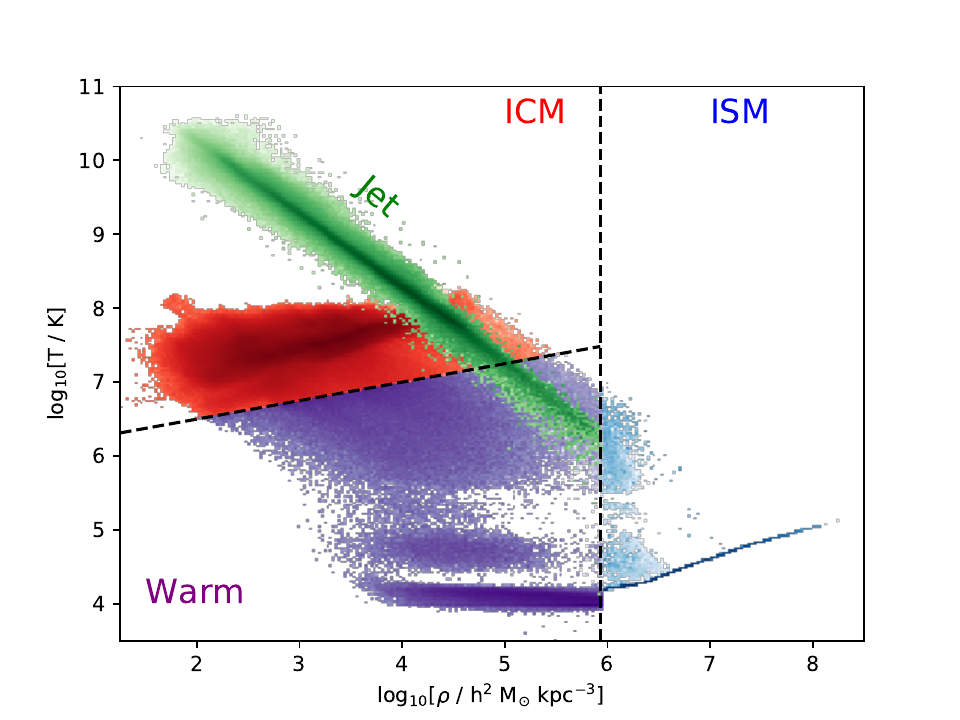,width=0.5\textwidth,angle=0}
\caption{Phase diagram of the fiducial jet run at $19.8$~Myr to illustrate how we define different jet phases. Jet material is shown in green, ISM ($\rho>\rho_{\rm crit}=8.5\times 10^{5}$~h$^{2}$~M$_{\odot}$~kpc$^{-3}$) in blue, ICM ($T_{\rm keV} > 3\times~10^6\rho^{0.25}_{\rm cgs}$) in red and the warm phase, which sits between the ISM and ICM, in purple.}
\label{fig:phases}
\end{figure}

In Section~\ref{sec:cluster_properties} we wish to examine the impact that the jet feedback has on different gas phases and how gas transitions between them. Hence, in Fig.~\ref{fig:phases} we show the phase diagram of the fiducial jet run at $19.8$~Myr which helps us identify four key phases, whereby the gas is split into jet material, which we have already defined as non-ISM gas with $f_{\rm J}>f_{\rm J}^{\rm thresh}$ (green), ISM,  which is any gas with densities above $\rho_{\rm crit}=8.5\times 10^{5}$~h$^{2}$~M$_{\odot}$~kpc$^{-3}$ and is considered star-forming (blue), the ICM, which is defined using the temperature-density cut, $T_{\rm keV} > 3\times~10^6\rho^{0.25}_{\rm cgs}$, of \citet{RasiaEtAl12} (red), and finally the warm phase, which sits between the ICM and ISM phases and is defined as $T_{\rm keV} < 3\times~10^6\rho^{0.25}_{\rm cgs}$ and $\rho < \rho_{\rm crit}$ (purple). 

\section{Computational resources}
\label{app:resources}
\begin{table*}  
\centering
\begin{tabular}{|l|c|c|c|}
\hline
 & Pre-jet/[$10^{3}$ core-hrs] & Jet/[$10^{3}$ core-hrs] & Post-jet/[$10^{3}$ core-hrs] \\
Run & ($-6$ to $0$ Myr) & ($0$ to $20$ Myr) & ($20$ to $100$ Myr) \\
\hline
low & $0.11$ ($0.11$) & $4.99$ ($4.98$) & $21.8$ ($16.8$) \\
fiducial & $0.11$ ($0.11$) & $11.3$ ($11.2$) & $37.7$ ($26.3$) \\
high  & $0.11$ ($0.11$) & $30.7$ ($30.6$) & $82.2$ ($51.5$)\\
\hline
\end{tabular}
\caption{Summary of total (phase) core hours used for each jet simulation split into three main phases: {\it pre-jet phase}, covering the $\sim 6$~Myrs prior to the jet switching on during which the system relaxes and the central refinement kicks in, the {\it jet phase}, covering the $20$~Myrs that the jets are switched on, and the {\it post-jet} phase covering the $\sim 80$~Myrs after the jet action halts.}
\label{tab:resources}
\end{table*}  

The simulations presented in this work make use of super-Lagrangian refinement techniques that extend the simulations dynamic range. This can be computationally expensive and for the interest of readers we have added Table~\ref{tab:resources}, which outlines the core hours used by the jet simulations. All simulations were performed on $512$ cores and underwent a {\it pre-jet} relaxation phase while the central refinement region developed (see second column). The core hours used for the main {\it jet} phase, comprising the first $20$~Myr of the simulation proper, are shown in the third column. Increasing jet power has a direct impact on the core hours used for this phase (see bracketed values), given higher jet velocities and temperatures (requiring smaller timesteps) and larger lobes with more refined cells in simulations with higher jet powers. Finally, the jet power also directly impacts the {\it post-jet} phase times, with this again being a reflection of cells having higher temperatures and there being more high resolution cells in simulations with higher jet powers. Additionally, as the jet power increases the {\it jet} phase represents a larger fraction of the total simulation time, a reflection of the computational effort required to simulate very fast, low density jets and lobes. 

\label{lastpage}
\end{document}